%% file: main_arXiv.tex
\documentclass[lettersize,journal]{IEEEtran}
\usepackage{amsmath,amsfonts}
\usepackage{algorithmic}
\usepackage{array}
\usepackage[caption=false,font=normalsize,labelfont=sf,textfont=sf]{subfig}
\usepackage{textcomp}
\usepackage{stfloats}
\usepackage{url}
\usepackage{verbatim}
\usepackage{graphicx}
\usepackage{balance}
\usepackage{amsmath}
\usepackage{amssymb}
\usepackage{amsthm}
\usepackage{mathtools}
\usepackage{enumitem}
\usepackage{color}
\usepackage{cite}
\usepackage{hyperref}
\usepackage{algorithm}

\newtheorem{lemma}{\hspace{0pt}\bf Lemma}
\newtheorem{proposition}{\hspace{0pt}\bf Proposition}
\newtheorem{example}{\hspace{0pt}\bf Example}

\newtheorem{theorem}{\hspace{0pt}\bf Theorem}

\newtheorem{definition}{\hspace{0pt}\bf Definition}
\newcommand{\argmin}{\operatornamewithlimits{argmin}}

\begin{document}

\title{Stabilizing Filters for Graph Signal Generation}

% \author{\IEEEauthorblockN{Martin Schmidt, \textit{Graduate Student Member, IEEE,} and Gonzalo Mateos, \textit{Senior Member, IEEE}}
%%%%% For arXiv preprint version comment above and use this one
\author{\IEEEauthorblockN{Martin Schmidt and Gonzalo Mateos}
\thanks{This work was supported in part by NSF under Grant ECCS 2231036. Part of the results in this paper were submitted to the \textit{2027 ICASSP Conference}~\cite{schmidt2026stablefilters}. \emph{(Corresponding author: Gonzalo Mateos.)}

Martin Schmidt and Gonzalo Mateos are with the Dept. of Electrical and Computer Engineering, University of Rochester, Rochester, NY 14627, USA (e-mails: mschmi21@ur.rochester.edu; gmateosb@ece.rochester.edu).}}

% \markboth{IEEE Transactions on Signal Processing,~Vol.~XX, No.~X, October~2026}%
% {Schmidt \MakeLowercase{\textit{et al.}}: Stabilizing Filters for Graph Signal Generation}

\maketitle

%%%%%%%%%%%%%%%%%%%%%%%%%%%%%%%%%%%%%%%%%%%%%%%%%%%%%%%%%%
%
%       Abstract
%
%%%%%%%%%%%%%%%%%%%%%%%%%%%%%%%%%%%%%%%%%%%%%%%%%%%%%%%%%%

\begin{abstract}
Generating graph signals is an important task in machine learning, and incorporating the underlying graph into a generative model enables structurally-informed distribution modeling. However, said graph is often noisy or imperfectly estimated, and structural errors can affect the generative process and the resulting signal distribution. While recent work has established stability guarantees for graph neural ordinary differential equations, the robustness of graph-aware reference dynamics, as used in topological Schr\"odinger bridge models, remains largely unexplored. Here, we investigate the structural stability of continuous-time generative models whose dynamics combine a graph filter with a learned graph neural network. We first derive a Wasserstein stability bound that quantifies the effect of graph perturbations on the generated signal distributions. Informed by this analysis, we then propose a principled framework for designing graph filters that balance structural stability with induced signal smoothing. We characterize the optimal filter design and provide an explicit family of polynomial filters with provable stability guarantees. Experiments on synthetic graphs and real-world fMRI signals support our theoretical findings, illustrating the enhanced structural stability of the proposed filters while preserving generative performance.
\end{abstract}


%%%%%%%%%%%%%%%%%%%%%%%%%%%%%%%%%%%%%%%%%%%%%%%%%%%%%%%%%%
%
%       Section I: Introduction
%
%%%%%%%%%%%%%%%%%%%%%%%%%%%%%%%%%%%%%%%%%%%%%%%%%%%%%%%%%%

\section{Introduction}

% Paragraph 1: graph signal generation motivation
Finding a map transporting a given data distribution to another is a fundamental problem underlying generative modeling, with numerous applications in machine learning and signal processing; see e.g.,~\cite{Welling_Lu_Holdijk_2026,xie2026spmag}. Recent advances in continuous-time generative modeling~\cite{lipman2022flow, ho2020denoisingdiffusionprobabilisticmodels, chen2019neuralordinarydifferentialequations} have achieved remarkable success generating realistic images, audio and video. However, their application to irregular, non-Euclidean domains that can be naturally represented using graphs is still fairly underdeveloped. While considerable research has focused on generating graph topologies~\cite{liu2019graphnormalizingflows, niu2020permutationinvariantgraphgeneration, jo2022scorebasedgenerativemodelinggraphs, liu2023generativediffusionmodelsgraphs, vignac2023digress,chen_2026}, far less attention has been paid to \emph{generating graph signals}~\cite{zhu2024graphsignaldiffusionmodel, uslu2026generative, rozada2025graphawarediffusionsignalgeneration,yang2025topological, wyrwal2026topological}. These signals are ubiquitous in real-world networked systems, such as neural time courses over structural brain connectomes~\cite{schmidt2025connectome}, vehicular flows on transportation networks~\cite{yu2017spatio}, and resource (e.g., power) allocations in wireless networks~\cite{camargo2026scalabilitygraphneuralnetwork}. Learning to faithfully generate graph signals has broad applicability, ranging from synthetic data augmentation and implicit priors to realistic system simulations across various timely domains. \vspace{2pt}

% Paragraph 2: graph-aware reference dynamics
\noindent\textbf{Graph-aware reference dynamics.} Among the many stochastic processes that match a given pair of endpoint distributions, the dynamic Schr\"odinger bridge problem provides a principled solution that favors the process closest to a prescribed reference dynamics~\cite{de2021diffusion, chen2022likelihood, liu2024generalized}. In the classical formulation, the reference is typically chosen as a standard Wiener process. However, when transporting distributions of graph signals, it is prudent to infuse the underlying structure directly into the generative model. Adopting custom graph-aware reference dynamics (e.g., heat diffusion over the graph)~\cite{yang2025topological, wyrwal2026topological, rozada2025graphawarediffusionsignalgeneration} promotes trajectories that are consistent with the graph topology.\vspace{2pt}

% Paragraph 3: the stability problem
\noindent\textbf{Permutation equivariance and stability.} By design, these generative models depend on the underlying graph to parametrize both the reference dynamics and the learned vector field. In practice, however, the observed graph is only an imperfect reflection of the true complex system~\cite[Ch. 7]{kolaczyk2009book}. For instance, correlation networks inferred from nodal measurements %derived from finite-length time series, 
can vary significantly with the sample size~\cite{zalesky2012use}, leading to noisy graph representations~\cite{SI_SPMAG, dong_2019_learning}. Consequently, unavoidable graph perturbations can skew the entire generative trajectory and the final distribution of the generated signals. This motivates the study of \emph{structural stability}, which quantifies the sensitivity of generative model outputs to perturbations in the graph parameterizing the generative dynamics. 

For graph-structured data, permutation equivariance is a desired property to ensure that the generative dynamics natively respect graph-induced symmetries~\cite{bronstein2021geometric}. Furthermore, because evaluating robustness across graphs with arbitrary node indexing calls for stability notions modulo permutation~\cite{Gama_2020}, equivariance is a mathematical prerequisite for any meaningful structural stability analysis, rather than merely a useful prior. \vspace{2pt}

% Paragraph 4: The gap in the literautre
\noindent\textbf{Innovations in context.} While permutation equivariance, structural stability, and transferability have been rigorously established for graph neural networks (GNNs)~\cite{Gama_2020, ruiz2023transferabilitypropertiesgraphneural,kenlay2021interpretablestabilityboundsspectral, levie2021transferability, keriven2020convergence}, these properties have primarily been studied in the context of static, discriminative tasks. In contrast, continuous-time generative models evolve a state trajectory that maps one distribution into another. Recent efforts have begun to bridge this gap, establishing structural stability~\cite{schmidt_2026} and transferability~\cite{yan2026convergencesizetransferabilitycontinuousdepth} bounds for graph neural ordinary differential equations (ODEs), i.e., GNN-parametrized ODEs~\cite{poli2021graphneuralordinarydifferential}. However, while the stability of a learned vector field can be controlled via regularization during training~\cite[Sec. V]{schmidt_2026}, graph-aware reference dynamics rely on a prescribed graph filter~\cite{yang2025topological, wyrwal2026topological, rozada2025graphawarediffusionsignalgeneration}. This raises the question of how to properly design stable reference dynamics.

% Paragraph 5: contributions 
In this context, our significant contributions to the foundations of graph signal generation are twofold: (i) we characterize the structural stability of generative models driven by graph-aware reference dynamics, deriving Wasserstein bounds that elucidate how graph perturbations affect the generated distributions; and (ii) we introduce an optimization framework to design graph filters that respect the induced smoothing of heat diffusion, while boosting robustness to graph imperfections.

\noindent\textbf{Paper outline.} The remainder of the paper is organized as follows. Section \ref{sec:preliminaries} reviews the necessary background on graph signal processing (GSP), the structural stability of graph operators, and generative modeling with topological Schr\"odinger bridges; as needed to formally state the problem. Section \ref{sec:stability_properties} formulates the continuous-time generative dynamics, and derives Wasserstein stability bounds under structural perturbations. Leveraging these theoretical insights, Section \ref{sec:stable_filters} introduces our principled filter design framework to delineate relevant stability-smoothness tradeoffs. To this end, we formulate a semi-infinite linear program and derive explicit optimal polynomial filter constructions. Section \ref{sec:experiments} numerically corroborates the efficacy of our optimal filter design relative to the heat diffusion baseline, both in a synthetic test case and a study involving fMRI brain signals. Finally, Section \ref{sec:conclusion} concludes the paper by summarizing our findings, discussing the impact of the chosen perturbation model on the optimal design, and contextualizing the practical upshot of our bounds. All technical details and proofs are deferred to the appendices.

Relative to the conference precursor \cite{schmidt2026stablefilters}, here we consider stabilizing graph filters for generative modeling through a unified presentation along with full technical details (including extended discussions and full proofs), supported by comprehensive numerical experiments. Noteworthy novel contributions in this journal paper include: (i) the full proofs of all theoretical results; (ii) the direct optimization of the exact integral Lipschitz constant, which departs from the surrogate bound minimized in \cite{schmidt2026stablefilters}; (iii) a rigorous theoretical characterization of the optimal filter convergence, including an explicit analytical polynomial construction that approaches the lower bound at an exponential rate; and (iv) a formal guarantee that optimality is attainable by a filter of finite order.

%%%%%%%%%%%%%%%%%%%%%%%%%%%%%%%%%%%%%%%%%%%%%%%%%%%%%%%%%%
%
%       Section II: Preliminaries
%
%%%%%%%%%%%%%%%%%%%%%%%%%%%%%%%%%%%%%%%%%%%%%%%%%%%%%%%%%%

\section{Preliminaries and Problem Statement}\label{sec:preliminaries}

Here we review key GSP concepts and survey GNN stability results, followed by the fundamentals of Schr\"odinger bridges for graph-supported data. We close with a problem statement.

\subsection{Graph Signal Processing}

Let $\mathcal{G}= (\mathcal{V}, \mathcal{E}, \mathcal{W})$ be a known undirected graph with $N$ nodes, edge set $\mathcal{E}\subseteq \mathcal{V}\times \mathcal{V}$, and edge weight map $\mathcal{W}: \mathcal{E}\to \mathbb{R}$. Its structure is represented by a graph shift operator (GSO) $\mathbf{L}\in\mathbb{R}^{N\times N}$, e.g., the graph Laplacian. A graph signal $x:\mathcal{V}\to\mathbb{R}$ is represented by a vector $\mathbf{x}\in\mathbb{R}^N$, with $x_i$ denoting the signal value at node $i$. By the spectral theorem, the symmetric GSO admits the eigendecomposition $\mathbf{L}=\mathbf{V}\mathbf{\Lambda}\mathbf{V}^\top$, with eigenvalues $\{\lambda_i\}_{i=1}^N$ and orthonormal eigenvectors $\mathbf{V}$. Accordingly, the graph Fourier transform (GFT) of $\mathbf{x}$ is $\hat{\mathbf{x}}=\mathbf{V}^\top\mathbf{x}$~\cite{ortega2018graph}. A graph convolutional filter of order $P$ is defined as $\mathbf{F}(\mathbf{L})\coloneqq\sum_{p=0}^{P}\theta_p\mathbf{L}^p$, with coefficients $\{\theta_p\}$~\cite{sandryhaila2014tsp,isufi2024tsp}, and corresponding frequency response $f(\lambda)\coloneqq\sum_{p=0}^{P}\theta_p\lambda^p$. A GNN extends this construction by cascading compositions of graph filters with pointwise nonlinearities $\sigma(\cdot)$~\cite{Gama_2019, kipf2016semi, defferrard2016convolutional}. For an input $\mathbf{x}$, we denote by $u_{\boldsymbol{\theta}}(\mathbf{x};\mathbf{L})$ the output of a GNN with learnable parameters $\boldsymbol{\theta}$ and fixed graph support $\mathbf{L}$. For convenience, we henceforth use $\mathcal{A}(\mathbf{x};\mathbf{L})$ to denote a generic graph-supported operator, say a graph filter or a GNN acting on $\mathbf{x}$, and parameterized by the GSO $\mathbf{L}$.

\subsection{Stability of Graph Filters and GNNs}
\label{sec: stability-GNN}

Let $\mathcal{P}$ denote the set of $N \times N$ permutation matrices. A desirable cardinal property of graph-supported operators is \emph{permutation equivariance}~\cite{Gama_2020}, meaning
\begin{equation*}
\mathcal{A}(\mathbf{P}^\top\mathbf{x};
\mathbf{P}^\top\mathbf{L}\mathbf{P})
=
\mathbf{P}^\top
\mathcal{A}(\mathbf{x};\mathbf{L})
\end{equation*}
for any permutation $\mathbf{P}\in\mathcal{P}$. In words, this means that e.g., GNN outputs are independent of arbitrary node labelings.

To analyze the robustness of graph filters and GNNs to structural errors, we adopt a well-documented relative perturbation model~\cite{Gama_2020, luana_2021}. Given a nominal GSO $\mathbf{L}$ and a perturbed one $\widetilde{\mathbf{L}}$, we evaluate the symmetric relative error matrix $\mathbf{E} \in \mathcal{S}:=\{\mathbf{S}\in\mathbb{R}^{N\times N}: \mathbf{S}=\mathbf{S}^\top\}$ at the permutation $\mathbf{P}_0 \in \mathcal{P}$ that minimizes the error's spectral norm, i.e.,
\begin{align}
\label{eq: perturbation-model}
\{\mathbf{E}^\star, \mathbf{P}_0\}
{}={}&
\argmin_{\mathbf{E}\in\mathcal{S},\, \mathbf{P}\in\mathcal{P}}
\|\mathbf{E}\|_2 \\
{}&{}\text{s.t.} \quad
\mathbf{P}^{\top}\widetilde{\mathbf{L}}\mathbf{P}
=
\mathbf{L}
+
\frac{1}{2}
\left(
\mathbf{E}\mathbf{L}
+
\mathbf{L}\mathbf{E}
\right).\nonumber
\end{align}
Unlike absolute perturbations where edge corruptions are uniform, this relative perturbation model ensures that regions of $\mathcal{G}$ with weaker connectivity experience proportionally smaller changes compared to others with stronger links; see also \cite[Sec. IV-A]{luana_2021} for a detailed discussion further justifying the practical relevance of this widely adopted model.

We next introduce the integral Lipschitz condition associated with graph filter stability under relative perturbations.
\begin{definition}
\label{def: integral-lipschitz}
Given a graph filter $\mathbf{F}$ with frequency response $f:\mathcal{D}\to\mathbb{R}$, where $\mathcal{D}\subseteq\mathbb{R}$, we say the filter is integral Lipschitz if there exists a constant $C\geq0$ such that
\begin{equation*}
|f(\lambda_2)-f(\lambda_1)|
\leq
2C
\frac{|\lambda_2-\lambda_1|}
{|\lambda_1+\lambda_2|},
\end{equation*}
for all $\lambda_1,\lambda_2\in\mathcal{D}$ such that $\lambda_1+\lambda_2\neq0$.
\end{definition}

For a differentiable frequency response $f$, taking the limit as $\lambda_2\to\lambda_1=\lambda$ in Definition~\ref{def: integral-lipschitz} implies that $|\lambda f'(\lambda)|\leq C$. The smallest admissible constant is the so-termed integral Lipschitz constant of $f$, denoted $C_{\mathrm{IL}}(f)$, and is equivalently given by
\begin{equation}
\label{eq: CIL}
C_{\mathrm{IL}}(f)
\coloneqq
\sup_{\substack{
\lambda_1,\lambda_2\in\mathcal{D}\\
\lambda_1 \neq \lambda_2
}}
\frac{|\lambda_1+\lambda_2|}{2}
\frac{|f(\lambda_2)-f(\lambda_1)|}
{|\lambda_2-\lambda_1|}.
\end{equation}

Graph filters and GNNs are structurally stable under the relative perturbation model \eqref{eq: perturbation-model}. Specifically, for a relative perturbation satisfying $\|\mathbf{E}^\star\|_2\leq\varepsilon$ and its corresponding optimal permutation $\mathbf{P}_0$, the pointwise deviation of a graph-supported operator satisfies
\begin{equation*}
\left\|
\mathbf{P}_0^\top
\mathcal{A}(\mathbf{x};\widetilde{\mathbf{L}})
-
\mathcal{A}(\mathbf{P}_0^\top\mathbf{x};\mathbf{L})
\right\|
\leq
\left(
\Gamma\varepsilon
+
\mathcal{O}(\varepsilon^2)
\right)
\|\mathbf{x}\|,
\end{equation*}
for all $\mathbf{x}\in\mathbb{R}^N$. The structural stability constant $\Gamma>0$ dictates the sensitivity of the operator to structural noise. For a standalone graph filter $\mathbf{F}$ with frequency response $f$, the corresponding stability constant is
\begin{equation*}
\Gamma_F
=
C_{\mathrm{IL}}(f)
\left(
1+\delta\sqrt{N}
\right),
\end{equation*}
where $\delta\leq 8$ represents the eigenvector misalignment between $\mathbf{E}^{\star}$ and the nominal GSO $\mathbf{L}$~\cite[Thm. 2]{Gama_2020}. Crucially, the integral Lipschitz constant $C_{\mathrm{IL}}(f)$ is the sole factor in this bound dictated by the filter design, whereas $\delta$ depends entirely on the realized structural perturbation. For a GNN, however, $\Gamma$ also depends on the network's depth, width, and the uniform spectral bounds of its intermediate filters~\cite[Thm. 4]{Gama_2020}. We build on these foundations to analyze the propagation of structural errors through the continuous-time dynamics of the following graph-aware generative model.

\subsection{Topological Schr\"odinger Bridge}

Let $\mathfrak{M}$ denote the set of probability measures on the path space $\mathcal{C}([0,1],\mathbb{R}^N)$. For any path measure $\mathbb{P}\in\mathfrak{M}$, we denote by $\mathbb{P}_t$ its time-$t$ marginal distribution; i.e., if $\mathbf{x}=(\mathbf{x}_t)_{0\leq t\leq 1}\sim\mathbb{P}$, then $\mathbf{x}_t\sim\mathbb{P}_t$. For prescribed initial and terminal marginals $\mathbf{x}_0\sim p_0$ and $\mathbf{x}_1\sim p_1$, and given a reference path measure $\mathbb{Q}\in\mathfrak{M}$, the Schr\"odinger bridge problem~\cite{leonard2013survey} seeks
\begin{equation}
\label{eq: sb}
\mathbb{P}^{\star}
\in
\arg\min_{\mathbb{P}\in \mathfrak{M}}
\mathrm{KL}(\mathbb{P}\|\mathbb{Q})
\quad \text{s.t.} \quad
\mathbb{P}_0=p_0,\ \mathbb{P}_1=p_1,
\end{equation}
where $\mathrm{KL}(\mathbb{P}\|\mathbb{Q})$ denotes the Kullback--Leibler divergence between the path measures $\mathbb{P}$ and $\mathbb{Q}$.

For graph-supported data, the reference process can be chosen to explicitly incorporate the underlying topology~\cite{yang2025topological}. Given GSO $\mathbf{L}$,  consider the graph-aware reference diffusion
\begin{equation}
\label{eq: reference-SDE}
d\mathbf{y}_t
=
\left[
\mathbf{H}_t(\mathbf{L})\mathbf{y}_t
+
\boldsymbol{\alpha}_t
\right]dt
+
\mathbf{\Sigma}_t\,d\mathbf{w}_t,
\end{equation}
where $\mathbf{w}_t$ is a standard Brownian motion in $\mathbb{R}^N$, $\mathbf{\Sigma}_t\in\mathbb{R}^{N\times N}$ controls the diffusion, $\boldsymbol{\alpha}_t\in\mathbb{R}^N$ is a bias term, and $\mathbf{H}_t(\mathbf{L})\coloneqq\sum_{k=0}^{K}h_k(t)\mathbf{L}^k$ is a time-dependent graph filter of order $K$. The reference path measure $\mathbb{Q}$ is then the law induced by \eqref{eq: reference-SDE}.

Typically, samples are all we have from the endpoint distributions $p_0$ and $p_1$. %typically available only through samples. 
We consider generative dynamics governed by a stochastic differential equation (SDE) of the form
\begin{equation}
\label{eq: generative-SDE}
d\mathbf{x}_t
=
b_t(\mathbf{x}_t;\mathbf{L})dt
+
\mathbf{\Sigma}_t\,d\mathbf{w}_t,
\qquad
\mathbf{x}_0\sim p_0,
\end{equation}
where
\begin{equation}
\label{eq: generative-drift}
b_t(\mathbf{x};\mathbf{L})
\coloneqq
\mathbf{H}_t(\mathbf{L})\mathbf{x}
+
\boldsymbol{\alpha}_t
+
u_t^{\boldsymbol{\theta}}(\mathbf{x};\mathbf{L}).
\end{equation}
Here, $u_t^{\boldsymbol{\theta}}(\cdot;\mathbf{L})$ is a learned corrective vector field. It is trained so that the path law induced by \eqref{eq: generative-SDE} approximates the Schr\"odinger bridge $\mathbb{P}^{\star}$ in \eqref{eq: sb}. Specifically, we parameterize this time-dependent vector field as $u_t^{\boldsymbol{\theta}}(\mathbf{x};\mathbf{L}) \coloneqq u_{\boldsymbol{\theta}}(g(\mathbf{x},t);\mathbf{L})$, where $u_{\boldsymbol{\theta}}(\cdot;\mathbf{L})$ is a GNN and $g(\mathbf{x},t)$ is a continuous temporal conditioning map that incorporates the time variable $t \in [0,1]$ into the signal; see \cite[Sec. III]{schmidt_2026} for a detailed discussion. Different training procedures can be used to estimate this vector field from samples, including Schr\"odinger bridge matching and flow matching methods~\cite{yang2025topological, wyrwal2026topological}.

\subsection{Problem Formulation}

Given a trained generative model mapping the initial distribution $p_0$ to the target distribution $p_1$ via the dynamics in \eqref{eq: generative-SDE}--\eqref{eq: generative-drift}, our objective is to formally establish its robustness to graph perturbations. Let $\mathbf{L}$ denote the nominal GSO and $\widetilde{\mathbf{L}}$ be a perturbed GSO satisfying the relative perturbation model \eqref{eq: perturbation-model} bounded by $\|\mathbf{E}^{\star}\|_2\leq\varepsilon$. Since node indexing is arbitrary, we must compare the nominal and perturbed dynamics modulo permutations. Our twofold goal is to: (i) bound the Wasserstein distance between the resulting generated distributions in terms of $\varepsilon$; and (ii) extract actionable insights from said bound.

%%%%%%%%%%%%%%%%%%%%%%%%%%%%%%%%%%%%%%%%%%%%%%%%%%%%%%%%%%
%
%       Section III: Stability Properties
%
%%%%%%%%%%%%%%%%%%%%%%%%%%%%%%%%%%%%%%%%%%%%%%%%%%%%%%%%%%

\section{Stability Properties}\label{sec:stability_properties}

Let $\Phi_t(\mathbf{x}_0;\mathbf{L})$ denote the solution of \eqref{eq: generative-SDE} at time $t$ with initial condition $\mathbf{x}_0\sim p_0$. For a permutation $\mathbf{P}\in\mathcal{P}$, we define the time-$t$ marginal distributions of the permuted nominal and perturbed generative dynamics as
\begin{equation}
\label{eq: dynamics}
\Phi_t(\mathbf{P}^\top\mathbf{x}_0;\mathbf{L}) \sim \pi_t^{\mathbf{P}},\quad \mathbf{P}^\top\Phi_t(\mathbf{x}_0;\widetilde{\mathbf{L}})\sim \widetilde{\pi}_t^{\mathbf{P}}.
\end{equation}

To ensure the generative dynamics respect the graph's structural symmetries, the governing SDE must be permutation equivariant. The graph filter $\mathbf{H}_t(\mathbf{L})$ and the GNN $u_{\boldsymbol{\theta}}(\cdot;\mathbf{L})$ are permutation equivariant by design (cf. Section~\ref{sec: stability-GNN}). To extend this property to the time-dependent vector field $u_t^{\boldsymbol{\theta}}$, we assume that the temporal conditioning map $g$ is permutation equivariant, satisfying
\begin{equation*}
    g(\mathbf{P}^\top\mathbf{x},t)
    =
    \mathbf{P}^\top g(\mathbf{x},t),
    \qquad
    \text{for all }\mathbf{P}\in\mathcal{P}.
\end{equation*}
Together with the permutation equivariance of the GNN, this implies that the time-dependent vector field is itself permutation equivariant, namely,
\begin{equation*}
    u_t^{\boldsymbol{\theta}}
    (\mathbf{P}^\top\mathbf{x};
    \mathbf{P}^\top\mathbf{L}\mathbf{P})
    =
    \mathbf{P}^\top
    u_t^{\boldsymbol{\theta}}(\mathbf{x};\mathbf{L}),
    \qquad
    \text{for all }\mathbf{P}\in\mathcal{P},
\end{equation*}
which follows directly from
$u_t^{\boldsymbol{\theta}}(\mathbf{x};\mathbf{L})
\coloneqq
u_{\boldsymbol{\theta}}(g(\mathbf{x},t);\mathbf{L})$.
The assumed equivariance of $g$ is satisfied in practice by applying a temporal embedding (e.g., via node-wise addition or feature concatenation) uniformly across all nodes; see \cite[Sec.~III]{schmidt_2026}. Such ways of incorporating temporal embeddings are widely used in generative models; see e.g.,~\cite{song2021scorebasedgenerativemodelingstochastic,nichol2021improved}.

We further assume that, for all $\mathbf{P}\in\mathcal{P}$ and $t\in[0,1]$, the bias and diffusion satisfy
$\mathbf{P}^\top\boldsymbol{\alpha}_t=\boldsymbol{\alpha}_t$
(e.g., a uniform drift $\boldsymbol{\alpha}_t=c\mathbf{1}$) and
$\mathbf{P}^\top\mathbf{\Sigma}_t\mathbf{P}=\mathbf{\Sigma}_t$
(e.g., isotropic noise $\mathbf{\Sigma}_t=\sigma_t\mathbf{I}$).%for all $t\in[0,1]$.

Moreover, graph filters and GNNs are Lipschitz with respect to their input signals \cite[Prop.~3]{schmidt_2026}. Hence, for fixed $\mathbf{L}$, the drift $b_t(\cdot;\mathbf{L})$ is Lipschitz continuous, ensuring well-posedness and uniqueness of the generative SDE. Under these conditions, node relabeling commutes with the stochastic dynamics.

\begin{proposition}
\label{prop: equivariance}
For any initial distribution $\mathbf{x}_0 \sim p_0$, the probability laws of the generated trajectories of \eqref{eq: generative-SDE} satisfy 
\begin{equation*}
\mathbf{P}^\top\Phi_t(\mathbf{x}_0;\mathbf{L}) \overset{d}{=} \Phi_t(\mathbf{P}^\top\mathbf{x}_0;\mathbf{P}^\top\mathbf{L}\mathbf{P}),
\end{equation*}
for all $\mathbf{P} \in \mathcal{P}$ and $t \in [0,1]$, where $\smash{\overset{d}{=}}$ means equal in distribution.
\end{proposition}
\noindent\textit{Proof.} See Appendix \ref{proof: equivariance}.

To establish the stability of the generated distributions, we first formally introduce the structural stability and Lipschitz continuity constants characterizing the underlying drift vector field components in \eqref{eq: generative-drift}. As stated in Section \ref{sec: stability-GNN}, graph filters and GNNs are fundamentally stable operators. For a perturbation $\|\mathbf{E}^\star\|_2 \leq \varepsilon$ and its corresponding optimal permutation $\mathbf{P}_0$, we define the structural stability constants $\Gamma_H$ and $\Gamma_u$ such that the filter and GNN in \eqref{eq: generative-drift} satisfy
\begin{align}
    \|\mathbf{P}_0^\top \mathbf{H}_t(\widetilde{\mathbf{L}})\mathbf{x}-\mathbf{H}_t(\mathbf{L})\mathbf{P}_0^\top \mathbf{x}\| &{}\leq\big(\Gamma_H\varepsilon+\mathcal{O}(\varepsilon^2)\big)\|\mathbf{x}\|, \label{eq:stability-graph-filter}\\
    \|\mathbf{P}_0^\top u_t^{\boldsymbol{\theta}}(\mathbf{x};\widetilde{\mathbf{L}})-u_t^{\boldsymbol{\theta}}(\mathbf{P}_0^\top \mathbf{x};\mathbf{L})\| &{}\leq\big(\Gamma_u\varepsilon+\mathcal{O}(\varepsilon^2)\big)\|g(\mathbf{x}, t)\|,\label{eq:stability-graph-neural-network}
\end{align}
for all $\mathbf{x}\in\mathbb{R}^N$ and $t \in [0,1]$. By Lipschitz continuity, both operators also satisfy a one-sided Lipschitz condition. We denote their respective one-sided Lipschitz constants as $m_H$ and $m_u$, satisfying
\begin{align}
    \langle \mathbf{H}_t(\mathbf{L})(\mathbf{x}-\mathbf{y}),\mathbf{x}-\mathbf{y}\rangle &{}\leq m_H\|\mathbf{x}-\mathbf{y}\|^2, \label{eq:one-sided-filter}\\
    \langle u_t^{\boldsymbol{\theta}}(\mathbf{x};\mathbf{L})-u_t^{\boldsymbol{\theta}}(\mathbf{y};\mathbf{L}),\mathbf{x}-\mathbf{y}\rangle &{}\leq m_u\|\mathbf{x}-\mathbf{y}\|^2,\label{eq:one-sided-gnn}
\end{align}
for all $\mathbf{x},\mathbf{y}\in\mathbb{R}^N$ and $t \in [0,1]$, where $m_H$ and $m_u$ are chosen as uniform upper bounds for the nominal GSO $\mathbf{L}$ and the admissible perturbed GSOs $\widetilde{\mathbf{L}}$. Combining these properties yields the structural stability and continuity of the full drift vector field $b_t(\mathbf{x};\mathbf{L})$, as asserted in the following lemmata.

\begin{lemma}
\label{lem: comp-stability}
The vector field $b_t(\mathbf{x};\mathbf{L})$ in \eqref{eq: generative-drift} is structurally stable, i.e., for the optimal permutation $\mathbf{P}_0$ it holds that
\begin{equation*}
\begin{aligned}
    \|\mathbf{P}_0^\top b_t(\mathbf{x};\widetilde{\mathbf{L}})-b_t(\mathbf{P}_0^\top \mathbf{x};\mathbf{L})\| \leq \varepsilon \big(\Gamma_H \|\mathbf{x}\| + \Gamma_u \|g(\mathbf{x}, t)\|\big) \\
    + \mathcal{O}(\varepsilon^2)(\|\mathbf{x}\| + \|g(\mathbf{x}, t)\|),
\end{aligned}
\end{equation*}
for all $\mathbf{x}\in\mathbb{R}^N$ and $t \in [0,1]$.
\end{lemma}
\noindent\textit{Proof.} See Appendix \ref{proof: comp-stability}.

\begin{lemma}
\label{lem: comp-lipschitz}
The vector field $b_t(\mathbf{x};\mathbf{L})$ in \eqref{eq: generative-drift} is one-sided Lipschitz with combined constant $m_b=m_H+m_u$, i.e.,
\begin{equation*}
    \langle b_t(\mathbf{x};\mathbf{L})-b_t(\mathbf{y};\mathbf{L}),\mathbf{x}-\mathbf{y}\rangle\leq m_b\|\mathbf{x}-\mathbf{y}\|^2,
\end{equation*}
for all $\mathbf{x},\mathbf{y}\in\mathbb{R}^N$ and $t \in [0,1]$. The same bound holds for the perturbed GSO $\widetilde{\mathbf{L}}$.
\end{lemma}
\noindent\textit{Proof.} See Appendix \ref{proof: comp-lipschitz}.

Equipped with these properties, we establish the following stability result for the distributions generated by the SDE \eqref{eq: generative-SDE}.

\begin{theorem}
\label{thm: wasserstein-stability}
For any initial distribution $\mathbf{x}_0\sim p_0$, assume that the components of the drift $b_t(\mathbf{x};\mathbf{L})$ in \eqref{eq: generative-drift} are permutation equivariant, and that $b_t$ satisfies the structural stability and one-sided Lipschitz properties stated in Lemmata \ref{lem: comp-stability} and \ref{lem: comp-lipschitz}. Then, the Wasserstein-2 ($W_2$) distance between the probability laws generated by the nominal and perturbed dynamics (cf. \eqref{eq: dynamics}) satisfies
\begin{equation}\label{eq:W2_stability_bound}
\min_{\mathbf{P}\in\mathcal{P}}
W_2\left(\widetilde{\pi}_t^{\mathbf{P}},\pi_t^{\mathbf{P}}\right)
\leq
\Omega_t \big( \Gamma_H C_{H, t} + \Gamma_u C_{u, t} \big) \varepsilon
+ \mathcal{O}(\varepsilon^2),
\end{equation}
for all $t\in[0,1]$, where $\Omega_t \coloneqq \int_0^t e^{m_b(t-s)}ds$ and 
\begin{equation*}
\begin{aligned}
C_{H,t}
&\coloneqq
\max_{\mathbf{P}\in\mathcal{P}}
\sup_{s\in[0,t]}
\left(
\mathbb{E}\left[
\left\|
\Phi_s(\mathbf{P}^\top\mathbf{x}_0;\mathbf{L})
\right\|^2
\right]
\right)^{\frac{1}{2}},
\\
C_{u,t}
&\coloneqq
\max_{\mathbf{P}\in\mathcal{P}}
\sup_{s\in[0,t]}
\left(
\mathbb{E}\left[
\left\|
g\big(\Phi_s(\mathbf{P}^\top\mathbf{x}_0;\mathbf{L}),s\big)
\right\|^2
\right]
\right)^{\frac{1}{2}}.
\end{aligned}
\end{equation*}
\end{theorem}

\noindent\textit{Proof.} See Appendix \ref{proof: wassestein-stability}.

We find that stability is governed by three ingredients: (i) $\Gamma_H$ and $\Gamma_u$, which capture the inherent stability of the graph filter and GNN adopted for the parameterization; (ii) $\Omega_t$, which is exponential in the combined Lipschitz constant $m_b$ and dictates how the continuous dynamics amplify errors; and (iii) $C_{H, t}$ and $C_{u, t}$, the trajectory suprema across all initial condition permutations. Evaluating the bound in Theorem~\ref{thm: wasserstein-stability} at the terminal $t=1$ yields the stability of generated samples.

\section{Stabilizing Graph Filters}
\label{sec:stable_filters}

With an eye towards enhanced stability, the GNN constants $\{\Gamma_u, m_u\}$ can be regularized during training~\cite[Sec. V]{schmidt_2026} but the filter constants $\{\Gamma_H, m_H\}$ are predetermined by the chosen reference dynamics. We now specialize the general reference dynamics to a class amenable to filter synthesis. Let the GSO $\mathbf{L}$ be a normalized graph Laplacian with spectrum contained in $[0,2]$. We consider a degenerate zero-diffusion setting \cite{wyrwal2026topological}, with $\boldsymbol{\alpha}_t=\mathbf{0}$, $\mathbf{\Sigma}_t=\mathbf{0}$, and a time-independent graph filter
\begin{equation}\label{eq: graph-filter}
    \mathbf{H}(\mathbf{L})
    =
    \sum_{k=0}^{K} h_k \mathbf{L}^k.
\end{equation}
The reference dynamics in \eqref{eq: reference-SDE} therefore reduce to
\begin{equation}
\label{eq: deterministic-reference}
    d\mathbf{y}_t
    =
    \mathbf{H}(\mathbf{L})\mathbf{y}_t\,dt,
\end{equation}
for some initial condition $\mathbf{y}_0$.

\begin{example}[Graph Heat Diffusion]
A canonical choice is $\mathbf{H}(\mathbf{L})=-\kappa\mathbf{L}$, $\kappa>0$, which yields the graph heat equation
\begin{equation}
\label{eq: heat}
    d\mathbf{y}_t
    =
    -\kappa\mathbf{L}\mathbf{y}_t\,dt.
\end{equation}
These dynamics diffuse the signal over $\mathcal{G}$ and progressively promote spatial smoothness, making heat diffusion a natural model for a broad class of network processes~\cite{delvenne2015diffusion,nikitin2022non}.
\end{example}

Motivated by Theorem~\ref{thm: wasserstein-stability} and the success of the heat equation in graph signal generation~\cite{de2021diffusion, chen2022likelihood, liu2024generalized}, we aim to design a graph filter that preserves the smoothing properties of \eqref{eq: heat} while boosting stability against structural perturbations. 

\subsection{Stability Constraints}

Theorem~\ref{thm: wasserstein-stability} establishes that the filter constants $\{\Gamma_H, m_H\}$ affect the stability of the generative dynamics. First, we characterize the one-sided Lipschitz constant $m_H$.

\begin{proposition}
\label{prop: one-sided-lipschitz}
Let $\mathbf{H}(\mathbf{L})$ be a graph filter with frequency response $h(\lambda)$, and let $\{\lambda_i\}_{i=1}^N$ denote the eigenvalues of the GSO $\mathbf{L}$. Then, the one-sided Lipschitz constant of $\mathbf{H}(\mathbf{L})$ is
\begin{equation}\label{eq: one-sided-constant}
    m_H
    =
    \max_{i\in\{1,\ldots,N\}} h(\lambda_i).
\end{equation}
\end{proposition}
\noindent\textit{Proof.} See Appendix \ref{proof: one-sided-lipschitz}.

To rule out artificial stability improvements arising from uniform contraction effects (which act independently of the graph structure), we restrict our design space by setting $h_0=0$ in \eqref{eq: graph-filter}. Hence, the graph-frequency response takes the form
\begin{equation}\label{eq:filter-freq-response}
    h(\lambda)
    =
    \sum_{k=1}^{K} h_k \lambda^k,
\end{equation}
parameterized by the coefficients $\mathbf{h}\coloneqq[h_1,\ldots,h_K]^\top$. Because $h_0=0$, we have $h(0)=0$. Moreover, $\lambda = 0$ is an eigenvalue of the graph Laplacian, so the smallest achievable one-sided Lipschitz constant is $m_H=0$ [cf. \eqref{eq: one-sided-constant}]. For given graph $\mathcal{G}$, it would suffice to enforce $h(\lambda_i)\leq0$ at its eigenvalues. To obtain a graph-uniform guarantee valid for any normalized graph Laplacian, we instead impose
\begin{equation}\label{eq:admissibility-constraint}
    h(\lambda)\leq0,
    \qquad
    \forall \lambda\in[0,2].
\end{equation}
This ensures $m_H=0$ regardless of the graph's spectrum.

Second, we address the structural stability constant $\Gamma_H$. As discussed in Section~\ref{sec: stability-GNN}, $\Gamma_H$ is proportional to the filter's integral Lipschitz constant $C_{\mathrm{IL}}(h)$ in \eqref{eq: CIL}. Having $m_H=0$, our remaining objective is to minimize $C_{\mathrm{IL}}(h)$; see \eqref{eq: filter-design}. Note that this criterion differs from (and refines) prior work~\cite{schmidt2026stablefilters}, which minimized the surrogate bound $\max_{\lambda}|\lambda h'(\lambda)|$ rather than optimizing the exact integral Lipschitz constant directly.

\subsection{Smoothness Constraints}

To quantify the smoothness induced by the reference dynamics, we consider the Dirichlet energy of the terminal signal. The solution of \eqref{eq: deterministic-reference} at $t=1$ has graph-frequency representation
\begin{equation*}
    \hat{y}_{1,i}
    =
    e^{h(\lambda_i)}\hat{y}_{0,i},
    \qquad
    i=1,\ldots,N.
\end{equation*}
Its Dirichlet energy is therefore
\begin{equation*}
    \mathbf{y}_1^\top\mathbf{L}\mathbf{y}_1
    =
    \sum_{i=1}^{N}
    \lambda_i e^{2h(\lambda_i)}
    |\hat{y}_{0,i}|^2.
\end{equation*}
For $\|\mathbf{y}_0\|\leq1$, the worst-case Dirichlet energy satisfies
\begin{equation*}
    \mathbf{y}_1^\top\mathbf{L}\mathbf{y}_1
    \leq
    \max_{\lambda\in[0,2]}
    \lambda e^{2h(\lambda)}
\end{equation*}
and we therefore impose the smoothness constraint
\begin{equation*}
    \max_{\lambda\in[0,2]}
    \lambda e^{2h(\lambda)}
    \leq
    \eta,
\end{equation*}
for a prescribed tolerance $\eta\in(0,2)$. Taking logarithms, this translates to an explicit requirement on the filter's graph-frequency response, namely
\begin{equation}\label{eq: smoothness-constraint}
    h(\lambda)
    \leq
    \frac{1}{2}
    \log\!\left(\frac{\eta}{\lambda}\right),
    \qquad
    \text{for all } \lambda\in(0,2].
\end{equation}
\subsection{Stabilizing Filter Design Problem}

\begin{figure}[t]
    \centering
    \includegraphics[width=\columnwidth]{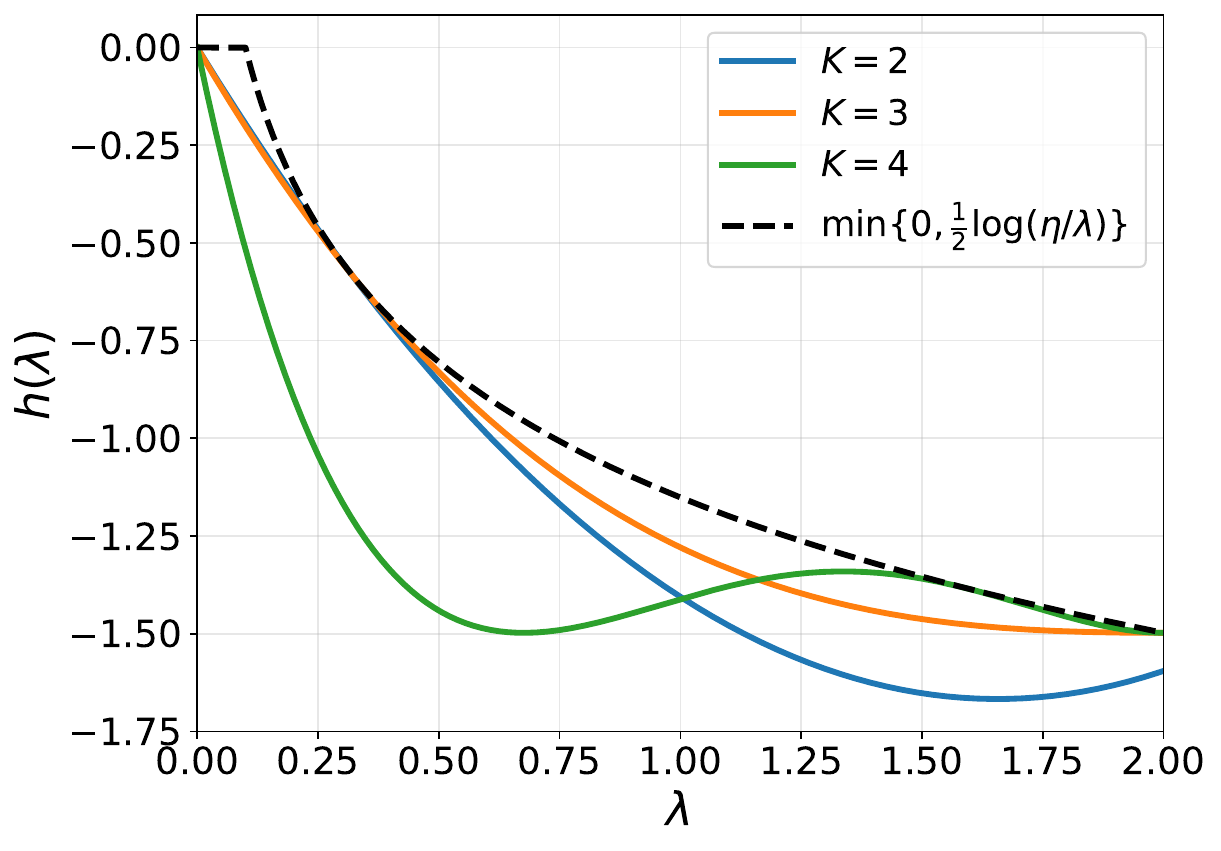}
    \caption{Numerical solutions to  \eqref{eq: filter-silp} for $\eta = 0.1$ and different polynomial orders $K$, together with the smoothness bound.}
    \label{fig:optimal-filters}
\end{figure}

Combining the stability and smoothness desiderata, we seek filter coefficients $\mathbf{h}\in\mathbb{R}^K$ that minimize the integral Lipschitz constant $C_{\mathrm{IL}}(h)$ while satisfying the admissibility and smoothness constraints in \eqref{eq:admissibility-constraint} and \eqref{eq: smoothness-constraint}, respectively. This gives the proposed stabilizing filter design problem
\begin{equation}
\label{eq: filter-design}
\begin{aligned}
\min_{\mathbf{h}}\quad
& C_{\mathrm{IL}}(h)
\\
\mathrm{s.t.}\quad
&
h(\lambda)
\leq
\min\left\{
0,
\frac{1}{2}\log\left(\frac{\eta}{\lambda}\right)\right\},
\qquad
\lambda\in(0,2].
\end{aligned}
\end{equation}
Using the definition of $C_{\mathrm{IL}}(h)$ in \eqref{eq: CIL}, we can equivalently write \eqref{eq: filter-design} in its epigraph form as
\begin{align}
\label{eq: filter-silp}
\min_{\mathbf{h},C}\quad
& C
\\
\mathrm{s.t.}\quad
&
|h(\lambda_2)-h(\lambda_1)|
\leq
2C\frac{\lambda_2-\lambda_1}{\lambda_1+\lambda_2},
\quad 0\leq\lambda_1<\lambda_2\leq2, \nonumber\\
&
h(\lambda)
\leq
\min\left\{
0,
\frac{1}{2}\log\!\left(\frac{\eta}{\lambda}\right)
\right\},
\quad
\lambda\in(0,2].\nonumber
\end{align}
We denote the optimal value of \eqref{eq: filter-silp} by $C_K^\star$. Since $h(\lambda)$ in \eqref{eq:filter-freq-response} is linear in the polynomial coefficients $\mathbf{h}$, \eqref{eq: filter-silp} is a semi-infinite linear program. In practice, it can be approximated by discretizing \((0,2]\) over a dense grid and solving the resulting finite-dimensional linear program \cite[Sec. 7]{hettich1993semi}. Representative numerically optimized filter responses for different polynomial orders are shown in Fig.~\ref{fig:optimal-filters}.

\subsection{Optimal Value and Explicit Constructions}

We will now characterize the convergence of \eqref{eq: filter-silp} to a fundamental limit, as the filter order $K$ grows. First, we observe that a lower bound on the optimal value $C_K^\star$ can be shown to be 
\begin{equation}\label{eq:lower-bound}
    C_K^\star \geq \frac{a}{2}, \quad \text{where } a =\frac{1}{2}\ln\left(\frac{2}{\eta}\right).
\end{equation}
This follows by evaluating the integral Lipschitz quotient at the endpoints $(\lambda_1, \lambda_2) = (0,2)$; see also Proposition~\ref{prop: CIL-convergence} for further details. Notice that the bound is independent of $K$.

Consider now the degree-$K$ family of filters with response
\begin{equation}
\label{eq: candidate-hKb}
h_{K,b_K}(\lambda)
\coloneqq
-b_K\left[
1-\left(1-\frac{\lambda}{2}\right)^K
\right],
\end{equation}
where
\begin{equation}
\label{eq: b-hat}
b_K
\coloneqq
\max_{x\in[\eta/2,1]}
\frac{
a+\frac{1}{2}\ln x
}{
1-(1-x)^K
}.
\end{equation}

\begin{figure}[t]
    \centering
    \includegraphics[width=\columnwidth]{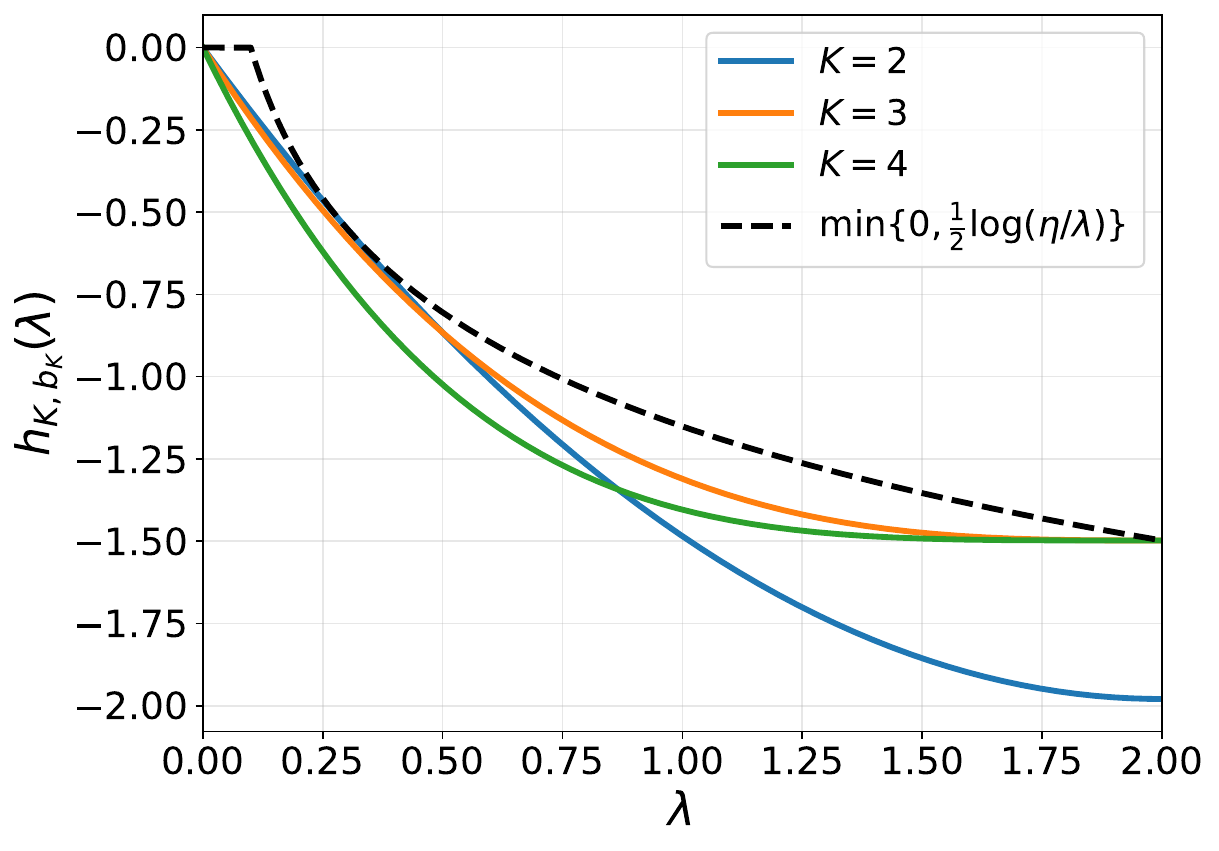}
    \caption{Filters $h_{K,b_K}$ in \eqref{eq: candidate-hKb} for $\eta = 0.1$ and different polynomial orders $K$, together with the smoothness bound. The claimed feasibility is apparent.}
    \label{fig:candidate-filters}
\end{figure}

Proposition~\ref{prop: CIL-convergence} asserts that these polynomials satisfy the constraints of \eqref{eq: filter-silp} for all $K\geq 2$, and that their integral Lipschitz constants approach the lower bound \eqref{eq:lower-bound} at a rate of $\mathcal{O}(r^K)$, where $r = 1 - \frac{\eta}{2}$. This provides an upper bound for the convergence of $C_K^\star$ and an explicit polynomial construction that approaches the theoretical optimum at an exponential rate; representative responses are shown in Fig.~\ref{fig:candidate-filters}.

\begin{proposition}
\label{prop: CIL-convergence}
For every $K\geq2$, let $h_{K,b_K}$ be given by \eqref{eq: candidate-hKb}. Then $h_{K,b_K}$ is feasible for \eqref{eq: filter-silp}, and
\begin{equation*}
\frac{a}{2}
\leq
C_K^\star
\leq
C_{\mathrm{IL}}\bigl(h_{K,b_K}\bigr)
\leq
\frac{a}{2(1-r^K)}.
\end{equation*}
Hence,
\begin{equation*}
C_K^\star
\longrightarrow
\frac{a}{2},
\qquad
C_K^\star-\frac{a}{2}
=
\mathcal{O}(r^K).
\end{equation*}
\end{proposition}
\noindent\textit{Proof.} See Appendix \ref{proof: convergence}.

Proposition~\ref{prop: CIL-convergence} established that the optimal integral Lipschitz constant approaches the universal lower bound exponentially with the polynomial order. The lower bound, however, need not only be approached asymptotically. A finite-order polynomial can attain it exactly, as the next result shows.

\begin{proposition}
\label{prop: finite-order-optimality}
Let $h_{K,b_K}$ be given by \eqref{eq: candidate-hKb}. If
\begin{equation*}
K
\geq
K_0
\coloneqq
\max\left\{
2,\,
1+
\left\lceil
\frac{\ln(2a)}{-\ln r}
\right\rceil
\right\},
\end{equation*}
then $b_K=a$, and 
\begin{equation*}
C_{\mathrm{IL}}\bigl(h_{K,b_K}\bigr)
=
C_K^\star
=
\frac{a}{2}.
\end{equation*}
\end{proposition}
\noindent\textit{Proof.} See Appendix \ref{proof: finite-order-optimality}.

The threshold $K_0$ is only a sufficient condition and is generally conservative. In practice, the optimum may be attained at substantially smaller orders. The main significance of Proposition \ref{prop: finite-order-optimality} is that the optimal value $a/2$ can always be attained by a finite-order polynomial.

\begin{figure*}[!t]
\centering
\includegraphics[width=0.49\textwidth]{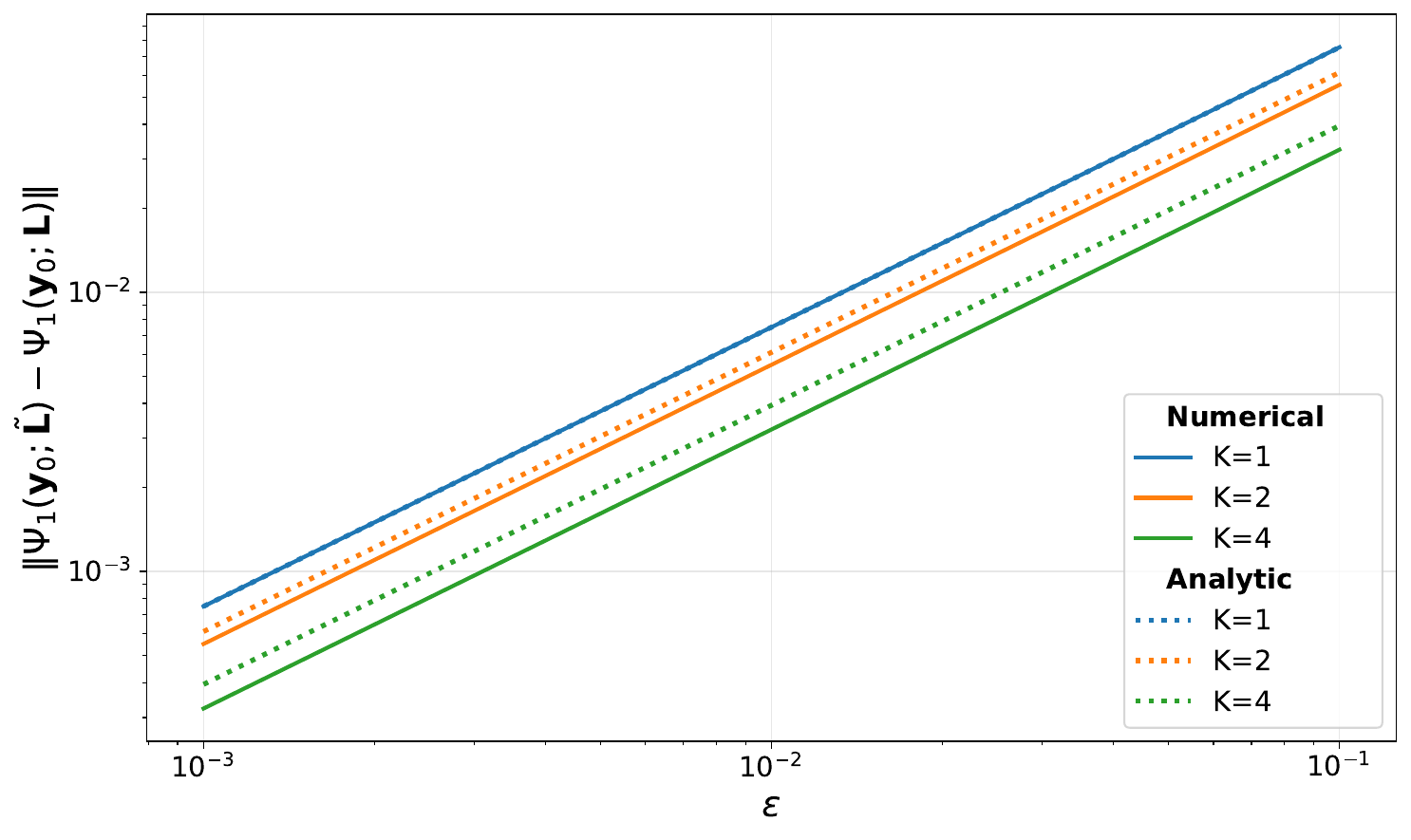}
\hfill
\includegraphics[width=0.49\textwidth]{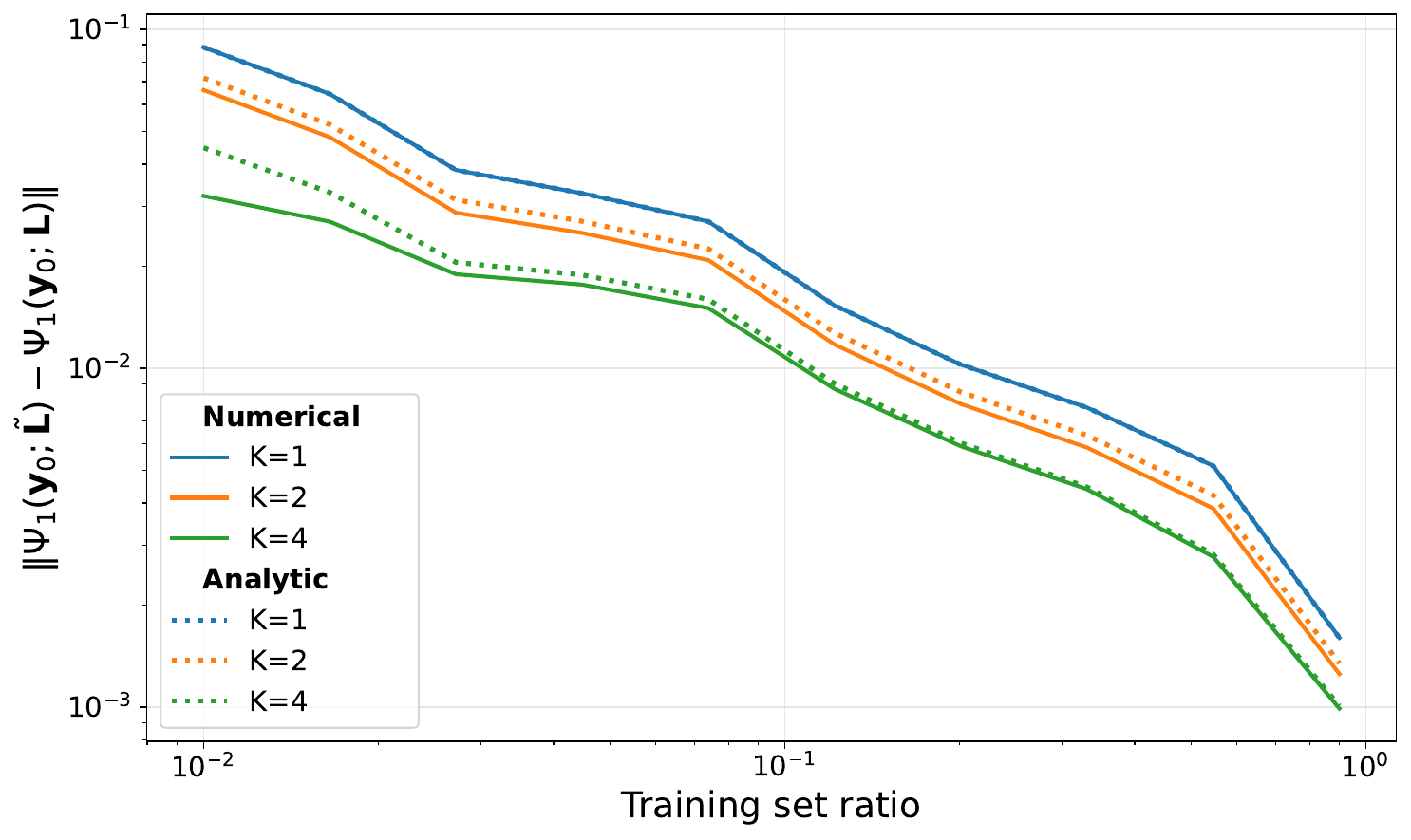}

\caption{Structural stability of the deterministic reference dynamics \eqref{eq: experimental-reference}. (Left) SBM graph under controlled synthetic perturbations of magnitude $\varepsilon$, and (right) fMRI graph constructed using different training set ratios. Solid lines correspond to the numerically optimized filters obtained from \eqref{eq: filter-silp}, while dotted lines represent the analytical construction in \eqref{eq: candidate-hKb}, for $\eta =0.1$. In both settings, structural stability improves with increasing polynomial order $K$, with the analytical filters exhibiting behavior comparable to their numerical counterparts.}
\label{fig:ref_dynm}

\end{figure*}

%%%%%%%%%%%%%%%%%%%%%%%%%%%%%%%%%%%%%%%%%%%%%%%%%%%%%%%%%%
%
%   Section VI: Experiments
%
%%%%%%%%%%%%%%%%%%%%%%%%%%%%%%%%%%%%%%%%%%%%%%%%%%%%%%%%%%
\section{Numerical Experiments}\label{sec:experiments}

We test the proposed stabilizing filters in two settings: (i) a synthetic setting based on a Stochastic Block Model (SBM) graph~\cite{holland1983stochastic}; and (ii) a real-world scenario using fMRI data~\cite{HCP}, where functional brain connectivity defines the graph structure. In the synthetic test case, the objective is data generation, transporting signals from an initial Gaussian distribution to a target data distribution. In the fMRI setting, the objective is distribution matching, transporting signals from a source data distribution (liberal) to a target distribution (aligned). 

For both settings, we consider two experiments. First, we investigate the structural stability of the reference dynamics
\begin{equation}
\label{eq: experimental-reference}
d\mathbf{y}_t
=
\mathbf{H}(\mathbf{L})\mathbf{y}_t\,dt,
\qquad
\mathbf{y}_0\sim p_0,
\end{equation}
whose solution at time $t$ is denoted by $\Psi_t(\mathbf{y}_0;\mathbf{L})$, comparing the numerical solution of \eqref{eq: filter-silp}
with the analytical construction in \eqref{eq: candidate-hKb}. Second, we evaluate the full generative dynamics
\begin{equation}
\label{eq: experimental-generative}
d\mathbf{x}_t
=
\left[
\mathbf{H}(\mathbf{L})\mathbf{x}_t
+
u_t^{\boldsymbol{\theta}}(\mathbf{x}_t;\mathbf{L})
\right]dt,
\qquad
\mathbf{x}_0\sim p_0,
\end{equation}
whose solution at time $t$ is denoted by $\Phi_t(\mathbf{x}_0;\mathbf{L})$, assessing both generative performance and structural
stability under perturbations of the underlying graph with nominal GSO $\mathbf{L}$.

\subsection{Experimental Setup}

\noindent\textbf{Datasets.}
In the synthetic setting, we use an SBM graph \cite{holland1983stochastic} with two communities of $10$ nodes each ($N=20$). The intra- and inter-community edge formation probabilities are $0.8$ and $0.1$, respectively. Signals in each community are drawn from a Gaussian distribution with standard deviation 1. One community has mean 1, while the other has mean $-1$. We generate 150 signals for training and validation (80\% and 20\%, respectively) and an additional 150 signals for testing.

For the fMRI experiments, we consider a single subject from the HCP dataset \cite{HCP}. Data live in $\mathbb{R}^{360 \times 1190}$, where $N=360$ is the number of brain regions (nodes) and $1190$ is the length of temporal window. Each time point is treated as an individual graph signal $\mathbf{x}^{(i)} \in \mathbb{R}^{360}$, yielding a dataset $\{\mathbf{x}^{(1)}, \dots, \mathbf{x}^{(1190)}\}$. To construct the source and target distributions for the distribution matching task, we decompose the signals into high and low frequency components using graph spectral filtering. Specifically, we use the 30th eigenvalue of the original graph normalized Laplacian as the threshold, applying a high-pass filter to obtain the liberal signals (initial distribution) and a low-pass filter to obtain the aligned signals (final distribution). These signals remain fixed throughout the experiments. We split the dataset into 70\% for training, 10\% for validation, and 20\% for testing, constructing the nominal GSO $\mathbf{L}$ solely from training signals. \vspace{2pt}

\noindent\textbf{Graph perturbations and evaluation.} Graph perturbations are introduced exclusively during testing to evaluate structural stability, while training is performed using the original graphs. For the SBM graphs, we introduce a controlled synthetic relative perturbation of the form [cf. \eqref{eq: perturbation-model}]
\begin{equation*}
    \widetilde{\mathbf{L}} = \mathbf{L} + \frac{1}{2}(\mathbf{E}\mathbf{L} + \mathbf{L}\mathbf{E}), \quad \text{with } \|\mathbf{E}\|_2 = \varepsilon.
\end{equation*}
Here, $\mathbf{E}$ is obtained by sampling a random symmetric matrix $\mathbf{M}$ and rescaling it as $\mathbf{E}=\varepsilon\mathbf{M}/\|\mathbf{M}\|_2$, ensuring $\|\mathbf{E}\|_2=\varepsilon$. For the fMRI data, graph perturbations are obtained by computing empirical correlation matrices from different-sized subsets of the training data. We quantify structural stability by the variation in terminal outputs, namely, $\|\Psi_1(\mathbf{y}_0;\widetilde{\mathbf{L}}) -\Psi_1(\mathbf{y}_0;\mathbf{L})\|$ for the reference dynamics and $\|\Phi_1(\mathbf{x}_0;\widetilde{\mathbf{L}}) -\Phi_1(\mathbf{x}_0;\mathbf{L})\|$ for the complete generative model. For the latter, we additionally evaluate generative performance using the Wasserstein-1 ($W_1$) distance between the generated and target distributions. Trajectories are approximated using an Euler sampler with 100 steps. \vspace{2pt}

\noindent\textbf{Architecture.} We parameterize the vector field $u_t^{\boldsymbol{\theta}}(\mathbf{x}_t; \mathbf{L})$ using a three-layer GNN ($L=3$). The temporal conditioning function is defined as $g(\mathbf{x}, t) = [\mathbf{x}, \mathbf{1}\textbf{emb}(t)^\top]$, where $\textbf{emb}(t) \in \mathbb{R}^{128}$ is a standard 128-dimensional sinusoidal time embedding. For the synthetic SBM case, we use three hidden layers with widths $[64, 32, 16]$, whereas for the fMRI case, we use $[256, 64, 32]$. In both cases, the input consists of a 1-dimensional node feature and a 128-dimensional time embedding, resulting in an input feature dimension of $F_0 = 129$. The output layer has a single feature, $F_{\mathrm{out}} = 1$. Each graph convolutional layer utilizes $P+1=4$ filter taps and use the normalized graph Laplacian as the GSO. We employ SiLU as the pointwise nonlinearity $\sigma(\cdot)$. The smoothing tolerance is set to $\eta = 0.04$ for SBM and $\eta = 0.1$ for fMRI, based on preliminary experimentation. \vspace{2pt}

\noindent\textbf{Training.}
We train the learned vector field $u_t^{\boldsymbol{\theta}}(\mathbf{x}_t; \mathbf{L})$ using topological flow matching~\cite{wyrwal2026topological}, regressing against graph-aware conditional vector fields with an independent coupling as detailed in Algorithm~\ref{alg:topological-flow-matching}. We further incorporate a structural stability regularization term into the training objective to control the Lipschitz constant of the GNN, following~\cite[Sec. V]{schmidt_2026}. For completeness, the corresponding conditional paths and vector fields are derived in Appendix~\ref{app:flow-matching}.

\begin{algorithm}[t]
\caption{Topological Flow Matching}
\label{alg:topological-flow-matching}
\begin{algorithmic}[1]
\REQUIRE $\mathbf{L}=\mathbf{V}\boldsymbol{\Lambda}\mathbf{V}^{\top}$,
polynomial order $K$, smoothness tolerance $\eta$,
coupling $\gamma$, learning rate $\rho$, network $u_t^{\boldsymbol{\theta}}$

\STATE Obtain $\{h_k\}_{k=0}^{K}$ numerically via
\eqref{eq: filter-silp} or analytically via
\eqref{eq: candidate-hKb}--\eqref{eq: b-hat}

\STATE Compute $h_i=\sum_{k=0}^{K}h_k\lambda_i^k$,
$i=1,\ldots,N$

\WHILE{not converged}
    \STATE Sample $(\mathbf{y}_0,\mathbf{y}_1)\sim\gamma$,
    $t\sim\mathcal{U}[0,1]$

    \STATE GFT:
    $\hat{\mathbf{y}}_0=\mathbf{V}^{\top}\mathbf{y}_0$,
    $\hat{\mathbf{y}}_1=\mathbf{V}^{\top}\mathbf{y}_1$

    \STATE Compute conditional path $\hat{\mathbf{y}}_t$ via \eqref{eq:app-spectral-path}

    \STATE Compute conditional vector field $\hat{u}_t^{\mathbf{y}_0,\mathbf{y}_1}$ via \eqref{eq:app-spectral-control}

    \STATE Inverse GFT:
    $\mathbf{y}_t=\mathbf{V}\hat{\mathbf{y}}_t$,
    $u_t^{\mathbf{y}_0,\mathbf{y}_1}
    =\mathbf{V}\hat{u}_t^{\mathbf{y}_0,\mathbf{y}_1}$

    \STATE $\boldsymbol{\theta}\leftarrow
    \boldsymbol{\theta}
    -\rho\nabla_{\boldsymbol{\theta}}
    \left\|
    u_t^{\boldsymbol{\theta}}(\mathbf{y}_t;\mathbf{L})
    -u_t^{\mathbf{y}_0,\mathbf{y}_1}
    \right\|^2$
\ENDWHILE
\end{algorithmic}
\end{algorithm}

\begin{figure*}[!t]
\centering
% First row: SBM
\includegraphics[width=0.49\textwidth]{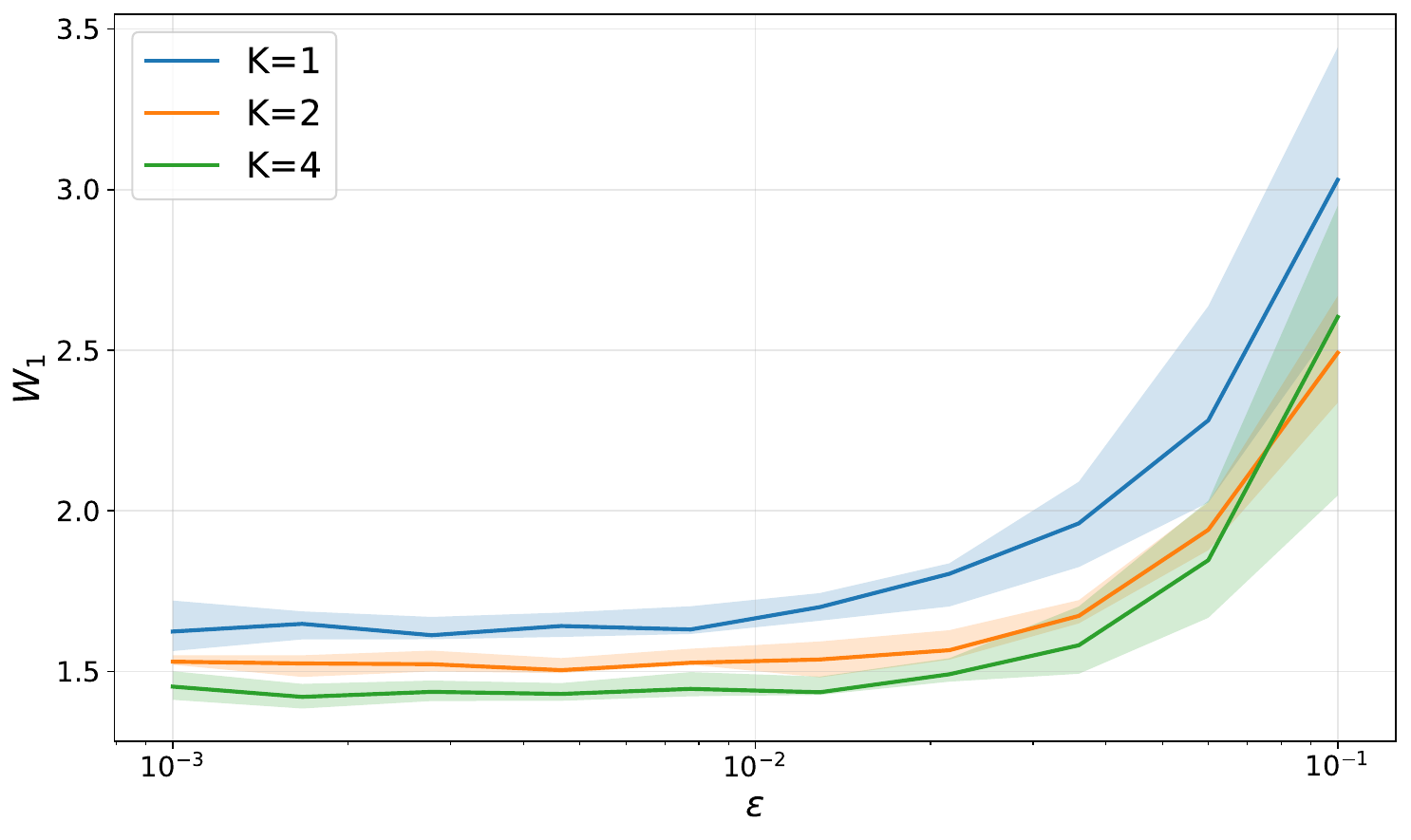}
\hfill
\includegraphics[width=0.49\textwidth]{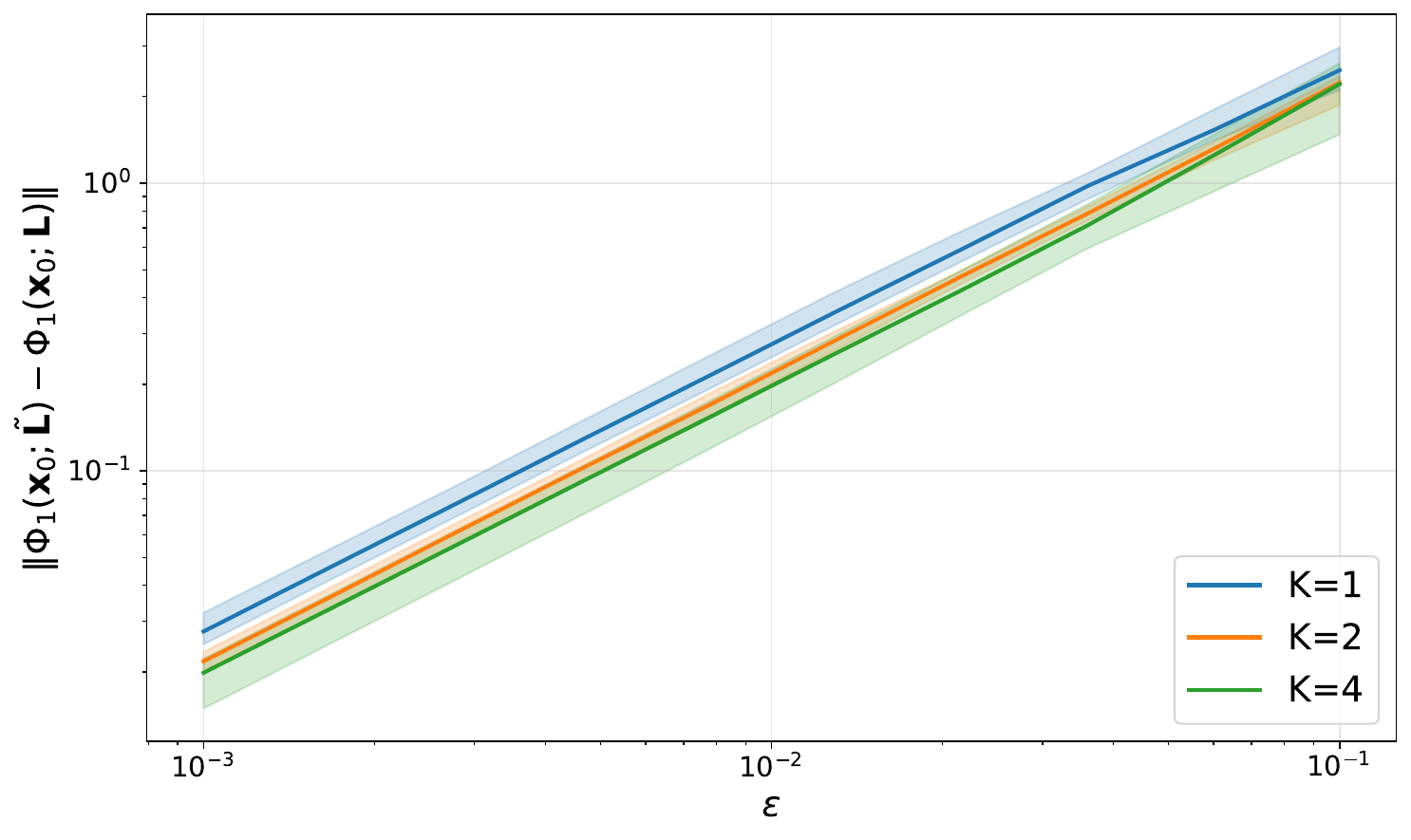}

\vspace{3pt} % Small vertical gap between rows

% Second row: fMRI
\includegraphics[width=0.49\textwidth]{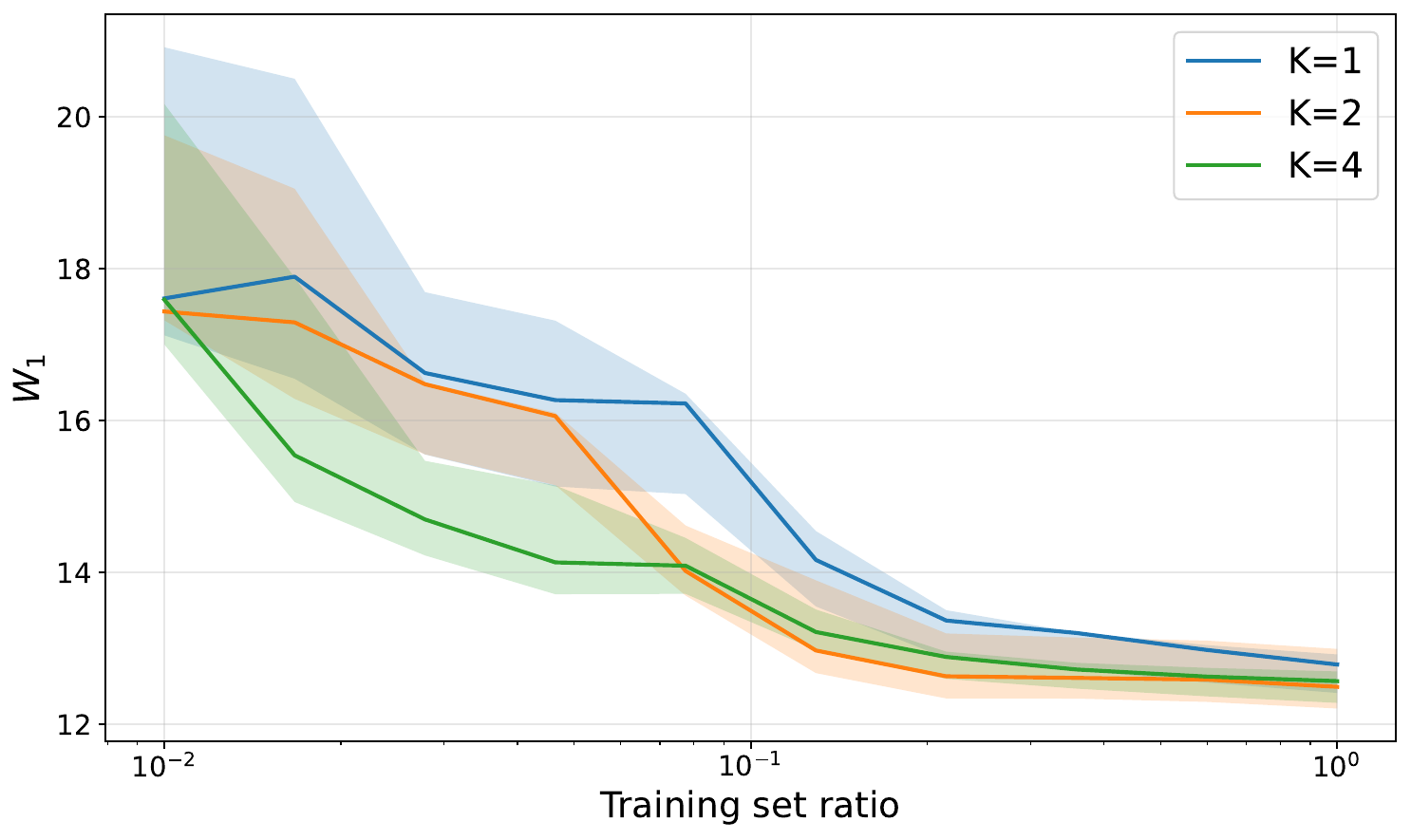}
\hfill
\includegraphics[width=0.49\textwidth]{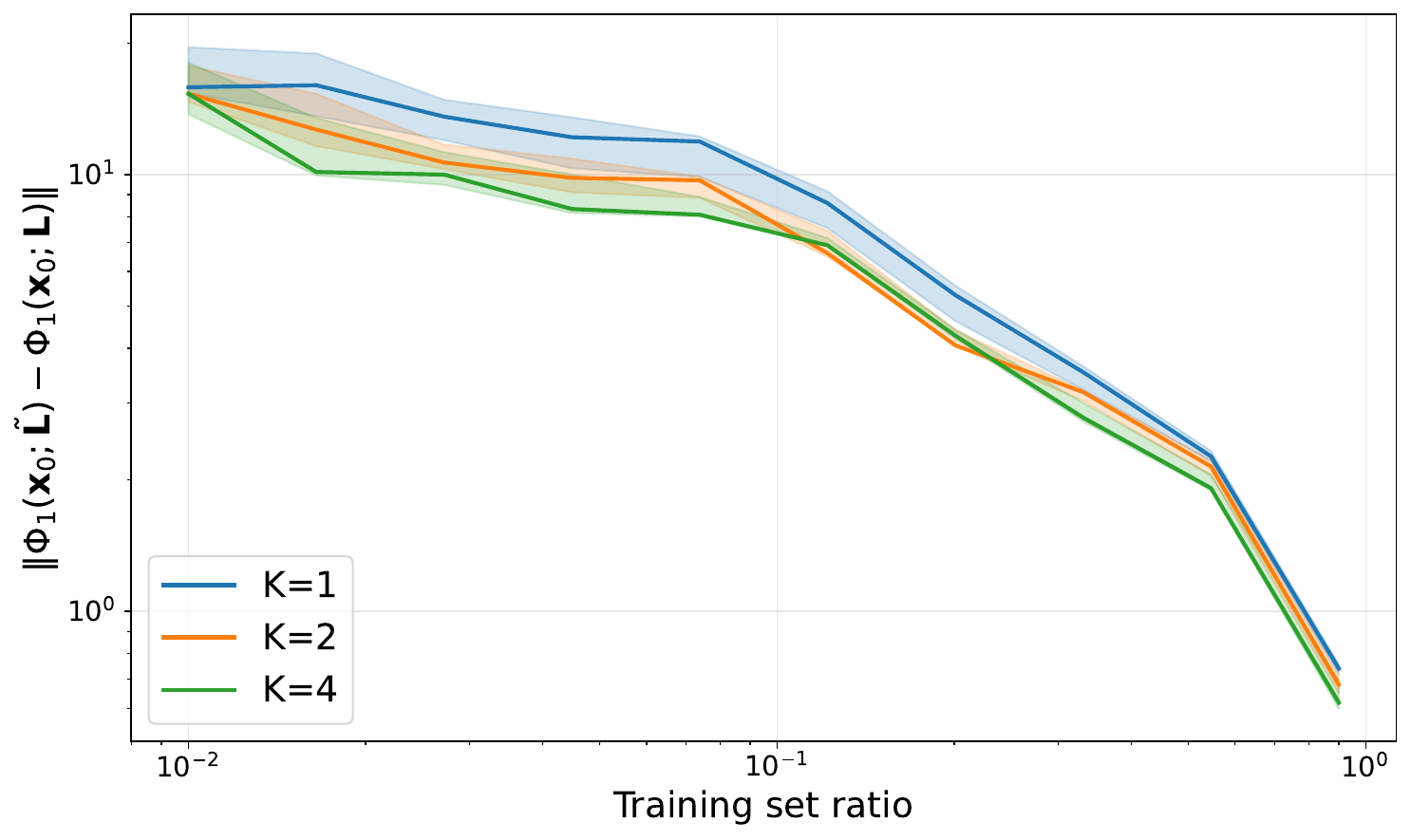}

\caption{Generative performance and structural stability of the generative dynamics \eqref{eq: experimental-generative} under graph perturbations. (Top) SBM graph under controlled synthetic perturbations of magnitude $\varepsilon$, and (bottom) fMRI graphs constructed using different training set ratios. (Left) Generative performance measured by the Wasserstein-1 ($W_1$) distance shows that higher-order optimized filters ($K=2,4$) achieve comparable or better performance than the heat equation baseline ($K=1$). (Right) Empirical stability measured by the variation in generated outputs under graph perturbations improves with increasing polynomial order $K$. Lines and shaded regions represent medians and 25th--75th percentiles across 10 independent runs.}
\label{fig:results}
\end{figure*}

\subsection{Results and Discussion}

\noindent\textbf{Reference dynamics.} Figure \ref{fig:ref_dynm} illustrates the structural stability of the isolated reference filters. As the polynomial order $K$ increases, the deviation between the terminal signals decreases, demonstrating improved robustness against graph perturbations. This trend holds for both the synthetic SBM graph (left) and the real-world fMRI (right). Also, the analytical filter construction in \eqref{eq: candidate-hKb} closely matches the stability performance of the numerically optimized filters in \eqref{eq: filter-silp}. \vspace{2pt}

\noindent\textbf{Generative model.} Figure \ref{fig:results} evaluates the full generative dynamics. While the learned GNN vector field $u_t^{\boldsymbol{\theta}}$ introduces an additional source of error, we still observe stability gains induced by the optimized graph filters. Specifically, higher-order filters ($K=2, 4$) exhibit lower output variation under perturbations compared to the heat equation baseline ($K=1$) across both datasets (right column). Importantly, the Wasserstein-1 distance metrics (left column) confirm that these structural stability improvements are achieved without degrading the performance of the generative model. Unlike the isolated reference dynamics, the stability of the complete generative model also depends on the learned GNN vector field, whose structural stability may vary across training runs. This additional variability makes the stability trends less pronounced, although the stabilizing effect of the reference filters generally carries over to the complete generative dynamics.

\section{Conclusion}\label{sec:conclusion}

This paper established a principled framework for designing stabilizing graph filters for graph-aware generative models. We demonstrate both theoretically and empirically that judiciously designed higher-order filters can improve structural stability over classical graph heat diffusion, while preserving generative performance. The novel graph filters can be obtained either by solving the semi-infinite linear program in \eqref{eq: filter-silp} for an optimal solution, or through the explicit polynomial construction in \eqref{eq: candidate-hKb}. For the latter, the parameter $b_K$ can be optimized for a prescribed smoothness level as in \eqref{eq: b-hat}, or else treated as a free design parameter.

An important consideration is that the proposed design is fundamentally tied to the relative perturbation model in \eqref{eq: perturbation-model}, which is well suited to graph-structured data \cite[Sec. IV-A]{luana_2021}. Different perturbation models lead to different stability bounds and, consequently, different optimal filter designs. For instance, under an absolute perturbation model, stability of GNNs is governed by the standard Lipschitz constant rather than the integral Lipschitz constant \cite[Thm. 1]{Gama_2020}, in which case the classical heat equation would be optimal.

Finally, although stability bounds may be numerically loose in practice \cite{schmidt_2026, schmidt2026stablefilters, Gama_2020}, their main value lies in their ability to expose the factors governing structural sensitivity. In this sense, they provide a principled foundation for both, the robust filter design developed in this work, and the regularization of the learned GNN vector field in~\cite{schmidt_2026}.

% --- Start Appendices ---
\appendices
\input{appendix}
% ------------------------

%\section*{Acknowledgments}
%Paste your acknowledgments here...

%\balance

%%%%% References %%%%%
\bibliographystyle{IEEEtran}
\bibliography{references}

%\begin{IEEEbiographynophoto}{First A. Author}
%Biography text here without a photo.
%\end{IEEEbiographynophoto}

% --- Uncomment below for a biography with a photo ---
% \begin{IEEEbiography}[{\includegraphics[width=1in,height=1.25in,clip,keepaspectratio]{author_photo.png}}]{Second B. Author}
% Biography text here with a photo.
% \end{IEEEbiography}

\end{document}

%% file: appendix.tex
%%%%%%%%%%%%%%%%%%%%%%%%%%%%%%%%%%%%%%%%%%%%%%%%%%%%%%%%%%
%
%                   Appendix A
%
%%%%%%%%%%%%%%%%%%%%%%%%%%%%%%%%%%%%%%%%%%%%%%%%%%%%%%%%%%

\section{Proof of Proposition \ref{prop: equivariance}}\label{proof: equivariance}
\begin{proof}
Let $\mathbf{z}_t \coloneqq \mathbf{P}^\top \mathbf{x}_t$. Then
\begin{equation*}
    d\mathbf{z}_t
    =
    \mathbf{P}^\top d\mathbf{x}_t
    =
    \mathbf{P}^\top b_t(\mathbf{x}_t;\mathbf{L})dt
    +
    \mathbf{P}^\top \mathbf{\Sigma}_td\mathbf{w}_t.
\end{equation*}
Since $b_t(\mathbf{x}_t;\mathbf{L})$ in \eqref{eq: generative-drift} is a sum of permutation-equivariant terms, we have
\begin{equation*}
    \mathbf{P}^\top b_t(\mathbf{x}_t;\mathbf{L})
    =
    b_t(\mathbf{P}^\top \mathbf{x}_t;\mathbf{P}^\top \mathbf{L}\mathbf{P})
    =
    b_t(\mathbf{z}_t;\mathbf{P}^\top \mathbf{L}\mathbf{P}).
\end{equation*}
Let $d\widetilde{\mathbf{w}}_t \coloneqq \mathbf{P}^\top d\mathbf{w}_t$. Since $\mathbf{P}$ is orthogonal, $\widetilde{\mathbf{w}}_t$ is also a standard Brownian motion. Moreover, since $\mathbf{P}^\top\mathbf{\Sigma}_t\mathbf{P} = \mathbf{\Sigma}_t$, we have
\begin{equation*}
    \mathbf{P}^\top \mathbf{\Sigma}_t\,d\mathbf{w}_t
    =
    \mathbf{\Sigma}_t \mathbf{P}^\top d\mathbf{w}_t
    =
    \mathbf{\Sigma}_t\,d\widetilde{\mathbf{w}}_t.
\end{equation*}
Therefore,
\begin{equation*}
    d\mathbf{z}_t
    =
    b_t(\mathbf{z}_t;\mathbf{P}^\top \mathbf{L}\mathbf{P})dt
    +
    \mathbf{\Sigma}_td\widetilde{\mathbf{w}}_t,
    \qquad
    \mathbf{z}_0 = \mathbf{P}^\top \mathbf{x}_0.
\end{equation*}
Thus, $\mathbf{z}_t$ satisfies the SDE associated with the initial condition $\mathbf{P}^\top \mathbf{x}_0$ and graph
$\mathbf{P}^\top \mathbf{L}\mathbf{P}$, driven by the standard Brownian motion $\widetilde{\mathbf{w}}_t$. Since $\widetilde{\mathbf{w}}_t$ and $\mathbf{w}_t$ have the same probability law, uniqueness in law of the SDE implies
\begin{equation*}
    \mathbf{P}^\top \Phi_t(\mathbf{x}_0;\mathbf{L})
    \overset{d}{=}
    \Phi_t(\mathbf{P}^\top \mathbf{x}_0;\mathbf{P}^\top \mathbf{L}\mathbf{P}),
\end{equation*}
which proves the result.
\end{proof}

%%%%%%%%%%%%%%%%%%%%%%%%%%%%%%%%%%%%%%%%%%%%%%%%%%%%%%%%%%
%
%                   Appendix B
%
%%%%%%%%%%%%%%%%%%%%%%%%%%%%%%%%%%%%%%%%%%%%%%%%%%%%%%%%%%

\section{Proof of Lemma \ref{lem: comp-stability}}\label{proof: comp-stability}
\begin{proof} From the definition of the vector field $b_t(\mathbf{x};\mathbf{L})$ in \eqref{eq: generative-drift} and leveraging the stability bounds for the graph filter \eqref{eq:stability-graph-filter} and the GNN \eqref{eq:stability-graph-neural-network}, a triangle inequality argument yields
\begin{equation*}
\begin{aligned}
&\|\mathbf{P}_0^\top b_t(\mathbf{x};\widetilde{\mathbf{L}})
-b_t(\mathbf{P}_0^\top\mathbf{x};\mathbf{L})\| \\
&\leq
\|\mathbf{P}_0^\top\mathbf{H}_t(\widetilde{\mathbf{L}})\mathbf{x}
-\mathbf{H}_t(\mathbf{L})\mathbf{P}_0^\top\mathbf{x}\|
\\ & +
\|\mathbf{P}_0^\top u_t^{\boldsymbol{\theta}}(\mathbf{x};\widetilde{\mathbf{L}})
-u_t^{\boldsymbol{\theta}}(\mathbf{P}_0^\top\mathbf{x};\mathbf{L})\| \\
&\leq
\varepsilon\big(
\Gamma_H\|\mathbf{x}\|
+
\Gamma_u\|g(\mathbf{x},t)\|
\big)
+
\mathcal{O}(\varepsilon^2)
\big(
\|\mathbf{x}\|+\|g(\mathbf{x},t)\|
\big).
\end{aligned}
\end{equation*}
\end{proof}

%%%%%%%%%%%%%%%%%%%%%%%%%%%%%%%%%%%%%%%%%%%%%%%%%%%%%%%%%%
%
%                   Appendix C
%
%%%%%%%%%%%%%%%%%%%%%%%%%%%%%%%%%%%%%%%%%%%%%%%%%%%%%%%%%%

\section{Proof of Lemma \ref{lem: comp-lipschitz}}\label{proof: comp-lipschitz} \begin{proof} From the definition of the vector field $b_t(\mathbf{x};\mathbf{L})$ in \eqref{eq: generative-drift} and recalling the one-sided Lipschitz constants for the graph filter \eqref{eq:one-sided-filter} and the GNN \eqref{eq:one-sided-gnn}, we have
\begin{equation*}
\begin{aligned}
&\langle
b_t(\mathbf{x};\mathbf{L})
-
b_t(\mathbf{y};\mathbf{L}),
\mathbf{x}-\mathbf{y}
\rangle \\
&=
\langle
\mathbf{H}_t(\mathbf{L})(\mathbf{x}-\mathbf{y}),
\mathbf{x}-\mathbf{y}
\rangle +
\langle
u_t^{\boldsymbol{\theta}}(\mathbf{x};\mathbf{L})
-
u_t^{\boldsymbol{\theta}}(\mathbf{y};\mathbf{L}),
\mathbf{x}-\mathbf{y}
\rangle \\
&\leq
(m_H+m_u)\|\mathbf{x}-\mathbf{y}\|^2.
\end{aligned}
\end{equation*}
\end{proof}

%%%%%%%%%%%%%%%%%%%%%%%%%%%%%%%%%%%%%%%%%%%%%%%%%%%%%%%%%%
%
%                   Appendix D
%
%%%%%%%%%%%%%%%%%%%%%%%%%%%%%%%%%%%%%%%%%%%%%%%%%%%%%%%%%%

\section{Proof of Theorem \ref{thm: wasserstein-stability}}\label{proof: wassestein-stability}
\begin{proof}
Fix the optimal permutation $\mathbf{P}_0 \in \mathcal{P}$ associated with the perturbation model in \eqref{eq: perturbation-model}. Let $\boldsymbol{\zeta} \sim p_0$ and let $\mathbf{w}_t$ be a standard Brownian motion. Define the nominal process $\mathbf{x}_t$ by
\begin{equation*}
d\mathbf{x}_t
=
b_t(\mathbf{x}_t;\mathbf{L})\,dt
+
\mathbf{\Sigma}_t\,d\mathbf{w}_t,
\qquad
\mathbf{x}_0
=
\mathbf{P}_0^\top \boldsymbol{\zeta}.
\end{equation*}
Likewise, define  the perturbed process $\widetilde{\mathbf{x}}_t$ by
\begin{equation*}
d\widetilde{\mathbf{x}}_t
=
b_t(\widetilde{\mathbf{x}}_t;\widetilde{\mathbf{L}})\,dt
+
\mathbf{\Sigma}_t\,d\widetilde{\mathbf{w}}_t,
\qquad
\widetilde{\mathbf{x}}_0
=
\boldsymbol{\zeta},
\end{equation*}
where $\widetilde{\mathbf{w}}_t
\coloneqq
\mathbf{P}_0 \mathbf{w}_t.$ Since $\mathbf{P}_0$ is orthogonal, $\widetilde{\mathbf{w}}_t$ is also a standard Brownian motion. Define $\mathbf{y}_t
\coloneqq
\mathbf{P}_0^\top \widetilde{\mathbf{x}}_t$ and notice that by construction,
\begin{equation*}
\mathbf{x}_t \sim \pi_t^{\mathbf{P}_0},
\qquad
\mathbf{y}_t \sim \widetilde{\pi}_t^{\mathbf{P}_0}.
\end{equation*}
Moreover,
\begin{align*}
d\mathbf{y}_t
&=
\mathbf{P}_0^\top d\widetilde{\mathbf{x}}_t \\
&=
\mathbf{P}_0^\top
b_t(\widetilde{\mathbf{x}}_t;\widetilde{\mathbf{L}})\,dt
+
\mathbf{P}_0^\top \mathbf{\Sigma}_t\,d\widetilde{\mathbf{w}}_t \\
&=
\mathbf{P}_0^\top
b_t(\mathbf{P}_0 \mathbf{y}_t;\widetilde{\mathbf{L}})\,dt
+
\mathbf{P}_0^\top \mathbf{\Sigma}_t \mathbf{P}_0\,d\mathbf{w}_t.
\end{align*}
Using $\mathbf{P}_0^\top \mathbf{\Sigma}_t \mathbf{P}_0 = \mathbf{\Sigma}_t$, we obtain
\begin{equation*}
d\mathbf{y}_t
=
\mathbf{P}_0^\top
b_t(\mathbf{P}_0 \mathbf{y}_t;\widetilde{\mathbf{L}})\,dt
+
\mathbf{\Sigma}_t\,d\mathbf{w}_t,
\qquad
\mathbf{y}_0
=
\mathbf{P}_0^\top \boldsymbol{\zeta}
=
\mathbf{x}_0.
\end{equation*}
Next, define $\mathbf{z}_t
\coloneqq
\mathbf{y}_t - \mathbf{x}_t.$ Since the two processes have the same initial condition and are driven by the same Brownian motion, the diffusion terms cancel, and therefore
\begin{equation*}
d\mathbf{z}_t
=
\left[
\mathbf{P}_0^\top
b_t(\widetilde{\mathbf{x}}_t;\widetilde{\mathbf{L}})
-
b_t(\mathbf{x}_t;\mathbf{L})
\right]dt,
\qquad
\mathbf{z}_0=\mathbf{0}.
\end{equation*}
We have
\begin{equation*}
\begin{aligned}
\frac{1}{2}\frac{d}{dt}\|\mathbf{z}_t\|^2
&=
\frac{1}{2}\frac{d}{dt}
\langle \mathbf{z}_t,\mathbf{z}_t\rangle \\
&=
\left\langle
\mathbf{z}_t,
\frac{d}{dt}\mathbf{z}_t
\right\rangle \\
&=
\left\langle
\mathbf{z}_t,
\mathbf{P}_0^\top
b_t(\widetilde{\mathbf{x}}_t;\widetilde{\mathbf{L}})
-
b_t(\mathbf{x}_t;\mathbf{L})
\right\rangle .
\end{aligned}
\end{equation*}
Adding and subtracting
$\mathbf{P}_0^\top
b_t(\mathbf{P}_0\mathbf{x}_t;\widetilde{\mathbf{L}})$
gives
\begin{equation*}
\begin{aligned}
\frac{1}{2}\frac{d}{dt}\|\mathbf{z}_t\|^2
&=
\left\langle
\mathbf{z}_t,
\mathbf{P}_0^\top
\left(
b_t(\widetilde{\mathbf{x}}_t;\widetilde{\mathbf{L}})
-
b_t(\mathbf{P}_0\mathbf{x}_t;\widetilde{\mathbf{L}})
\right)
\right\rangle
\\
&\quad
+
\left\langle
\mathbf{z}_t,
\mathbf{P}_0^\top
b_t(\mathbf{P}_0\mathbf{x}_t;\widetilde{\mathbf{L}})
-
b_t(\mathbf{x}_t;\mathbf{L})
\right\rangle .
\end{aligned}
\end{equation*}
We bound the two terms separately. For the first term, using the orthogonality of $\mathbf{P}_0$ and Lemma~\ref{lem: comp-lipschitz}, we have
\begin{equation*}
\begin{aligned}
&
\left\langle
\mathbf{z}_t,
\mathbf{P}_0^\top
\left(
b_t(\widetilde{\mathbf{x}}_t;\widetilde{\mathbf{L}})
-
b_t(\mathbf{P}_0\mathbf{x}_t;\widetilde{\mathbf{L}})
\right)
\right\rangle
\\
&\qquad =
\left\langle
\widetilde{\mathbf{x}}_t-\mathbf{P}_0\mathbf{x}_t,
b_t(\widetilde{\mathbf{x}}_t;\widetilde{\mathbf{L}})
-
b_t(\mathbf{P}_0\mathbf{x}_t;\widetilde{\mathbf{L}})
\right\rangle .
\\
&\qquad
\leq
m_b\|\widetilde{\mathbf{x}}_t-\mathbf{P}_0\mathbf{x}_t\|^2 = m_b \|\mathbf{z}_t\|^2 .
\end{aligned}
\end{equation*}
For the second term, the Cauchy--Schwarz inequality gives
\begin{equation*}
\begin{aligned}
&
\left\langle
\mathbf{z}_t,
\mathbf{P}_0^\top
b_t(\mathbf{P}_0\mathbf{x}_t;\widetilde{\mathbf{L}})
-
b_t(\mathbf{x}_t;\mathbf{L})
\right\rangle
\\
&\qquad \leq
\|\mathbf{z}_t\|
\left\|
\mathbf{P}_0^\top
b_t(\mathbf{P}_0\mathbf{x}_t;\widetilde{\mathbf{L}})
-
b_t(\mathbf{x}_t;\mathbf{L})
\right\|.
\end{aligned}
\end{equation*}
Applying Lemma~\ref{lem: comp-stability} with
$\mathbf{P}_0\mathbf{x}_t$ as the input yields
\begin{equation*}
\begin{aligned}
&
\left\|
\mathbf{P}_0^\top
b_t(\mathbf{P}_0\mathbf{x}_t;\widetilde{\mathbf{L}})
-
b_t(\mathbf{x}_t;\mathbf{L})
\right\|
\\
&\qquad \leq
\varepsilon
\left(
\Gamma_H\|\mathbf{P}_0\mathbf{x}_t\|
+
\Gamma_u
\|g(\mathbf{P}_0\mathbf{x}_t,t)\|
\right)
\\
&\qquad\quad
+
\mathcal{O}(\varepsilon^2)
\left(
\|\mathbf{P}_0\mathbf{x}_t\|
+
\|g(\mathbf{P}_0\mathbf{x}_t,t)\|
\right).
\end{aligned}
\end{equation*}
Since $\mathbf{P}_0$ is orthogonal and $g$ is permutation equivariant, then 
\begin{equation*}
\begin{aligned}
&
\left\langle
\mathbf{z}_t,
\mathbf{P}_0^\top
b_t(\mathbf{P}_0\mathbf{x}_t;\widetilde{\mathbf{L}})
-
b_t(\mathbf{x}_t;\mathbf{L})
\right\rangle \\
&\qquad \leq
\big(\varepsilon
\left(
\Gamma_H\|\mathbf{x}_t\|
+
\Gamma_u
\|g(\mathbf{x}_t,t)\|
\right)
\\
&\qquad\quad
+
\mathcal{O}(\varepsilon^2)
\left(
\|\mathbf{x}_t\|
+
\|g(\mathbf{x}_t,t)\|
\right)\big)\|\mathbf{z}_t\|.
\end{aligned}
\end{equation*}
Combining the two bounds, we obtain
\begin{equation*}
\begin{aligned}
\frac{1}{2}\frac{d}{dt}\|\mathbf{z}_t\|^2 \leq m_b \|\mathbf{z}_t\|^2 + \Big( \varepsilon \big(\Gamma_H \|\mathbf{x}_t\| + \Gamma_u \|g(\mathbf{x}_t, t)\|\big) \\
+ \mathcal{O}(\varepsilon^2)\big(\|\mathbf{x}_t\| + \|g(\mathbf{x}_t, t)\|\big) \Big) \|\mathbf{z}_t\| .
\end{aligned}
\end{equation*}
Let $r_t \coloneqq \|\mathbf{z}_t\|$, and by the chain rule one has $\frac{1}{2}\frac{d}{dt}r_t^2 =r_t\frac{d}{dt}r_t$. For $t$ such that $r_t>0$, dividing by $r_t$ gives
\begin{equation*}
\begin{aligned}
\frac{d}{dt}r_t
\leq
m_b r_t
+
\varepsilon
\big(
\Gamma_H\|\mathbf{x}_t\|
+
\Gamma_u\|g(\mathbf{x}_t,t)\|
\big)
\\
+
\mathcal{O}(\varepsilon^2)
\big(
\|\mathbf{x}_t\|
+
\|g(\mathbf{x}_t,t)\|
\big).
\end{aligned}
\end{equation*}
Let $a$ denote the left endpoint of the connected component of $\{s:r_s>0\}$ containing $t$. Since $r_0=0$, such a left endpoint exists, and by continuity $r_a=0$. Applying Gr\"onwall's inequality \footnote{If
$r'(t)\leq \mu r(t)+f(t)$ with $r(a)=0$, then
$r(t)\leq \int_a^t e^{\mu(t-s)}f(s)\,ds$.} on $[a,t]$, and noting that $[a,t]\subseteq[0,t]$ and the resulting integrand is nonnegative, we obtain
\begin{equation*}
\begin{aligned}
r_t
\leq
\int_0^t e^{m_b(t-s)}
\Big[
&\varepsilon
\big(
\Gamma_H\|\mathbf{x}_s\|
+
\Gamma_u\|g(\mathbf{x}_s,s)\|
\big)
\\
&+
\mathcal{O}(\varepsilon^2)
\big(
\|\mathbf{x}_s\|
+
\|g(\mathbf{x}_s,s)\|
\big)
\Big]\,ds.
\end{aligned}
\end{equation*}
The 2-Wasserstein distance is bounded by the cost of this coupling, namely
$W_2(\widetilde{\pi}_t^{\mathbf{P}_0}, \pi_t^{\mathbf{P}_0}) \leq \left(\mathbb{E}\big[\|\mathbf{z}_t\|^2\big]\right)^{\frac{1}{2}}$. Applying Minkowski's integral inequality \footnote{
$\left(\mathbb{E}\left[\left|\int_0^t f_s\,ds\right|^2\right]\right)^{\frac{1}{2}}
\leq
\int_0^t
\left(\mathbb{E}\left[|f_s|^2\right]\right)^{\frac{1}{2}}ds.$} and the $L^2$ triangle inequality \footnote{
$\left(\mathbb{E}\left[|X+Y|^2\right]\right)^{\frac{1}{2}}
\leq
\left(\mathbb{E}\left[|X|^2\right]\right)^{\frac{1}{2}}
+
\left(\mathbb{E}\left[|Y|^2\right]\right)^{\frac{1}{2}}.$}, yields
\begin{equation*}
\begin{aligned}
& \left(\mathbb{E}\big[\|\mathbf{z}_t\|^2\big]\right)^{\frac{1}{2}}
\leq 
\\
&
\varepsilon
\int_0^t e^{m_b(t-s)}
\Big(
\Gamma_H \left(\mathbb{E}\big[\|\mathbf{x}_s\|^2\big]\right)^{\frac{1}{2}}
+
\Gamma_u \left(\mathbb{E}\big[\|g(\mathbf{x}_s,s)\|^2\big]\right)^{\frac{1}{2}}
\Big)ds
\\
&
+
\mathcal{O}(\varepsilon^2)
\int_0^t e^{m_b(t-s)}
\Big(
\left(\mathbb{E}\big[\|\mathbf{x}_s\|^2\big]\right)^{\frac{1}{2}}
+
\left(\mathbb{E}\big[\|g(\mathbf{x}_s,s)\|^2\big]\right)^{\frac{1}{2}}
\Big)ds .
\end{aligned}
\end{equation*}
Using $C_{H,t} \geq \big(\mathbb{E}[\|\mathbf{x}_s\|^2]\big)^\frac{1}{2}$ and $C_{u,t} \geq \big(\mathbb{E}[\|g(\mathbf{x}_s,s)\|^2]\big)^\frac{1}{2}$ yields

\begin{equation*}
\begin{aligned}
& \min_{\mathbf{P}\in\mathcal{P}}
W_2\left(\widetilde{\pi}_t^{\mathbf{P}},\pi_t^{\mathbf{P}}\right)
\leq W_2\left(\widetilde{\pi}_t^{\mathbf{P}_0},\pi_t^{\mathbf{P}_0}\right) 
\\
& \qquad \leq
\Omega_t
\big(\Gamma_H C_{H,t} + \Gamma_u C_{u,t}\big)
\varepsilon
+
\mathcal{O}(\varepsilon^2)
\end{aligned}
\end{equation*}
which is the desired structural stability bound.
\end{proof}

%%%%%%%%%%%%%%%%%%%%%%%%%%%%%%%%%%%%%%%%%%%%%%%%%%%%%%%%%%
%
%                   Appendix E
%
%%%%%%%%%%%%%%%%%%%%%%%%%%%%%%%%%%%%%%%%%%%%%%%%%%%%%%%%%%

\section{Proof of Proposition \ref{prop: one-sided-lipschitz}}\label{proof: one-sided-lipschitz}
\begin{proof}
For any $\mathbf{x},\mathbf{y}\in\mathbb{R}^N$, because $\mathbf{H}(\mathbf{L})$ is symmetric,
\begin{equation*}
\begin{aligned}
&
\left\langle
\mathbf{H}(\mathbf{L})(\mathbf{x}-\mathbf{y}),
\mathbf{x}-\mathbf{y}
\right\rangle
\leq
\lambda_{\max}\big(\mathbf{H}(\mathbf{L})\big)
\|\mathbf{x}-\mathbf{y}\|^2.
\end{aligned}
\end{equation*}
Since the eigenvalues of $\mathbf{H}(\mathbf{L})$ are
$\{h(\lambda_i)\}_{i=1}^N$, then
\begin{equation*}
\lambda_{\max}\big(\mathbf{H}(\mathbf{L})\big)
=
\max_{i\in\{1,\ldots,N\}} h(\lambda_i)
\end{equation*}
and the result follows from the definition of one-sided Lipschitz constant. %Hence,
%
%\begin{equation*}
%m_H
%=
%\max_{i\in\{1,\ldots,N\}} h(\lambda_i).
%\end{equation*}
%
\end{proof}

%%%%%%%%%%%%%%%%%%%%%%%%%%%%%%%%%%%%%%%%%%%%%%%%%%%%%%%%%%
%
%                   Appendix F
%
%%%%%%%%%%%%%%%%%%%%%%%%%%%%%%%%%%%%%%%%%%%%%%%%%%%%%%%%%%

\section{Proof of Proposition \ref{prop: CIL-convergence}}\label{proof: convergence}
\begin{proof}
We first establish the lower bound. Evaluating the smoothness constraint at $\lambda=2$ gives
\begin{equation*}
    h(2)
    \leq
    \frac{1}{2}\ln\left(\frac{\eta}{2}\right)
    =
    -a.
\end{equation*}
Since $h(0)=0$, evaluating the integral Lipschitz quotient \eqref{eq: CIL} at
$(\lambda_1,\lambda_2)=(0,2)$ gives
\begin{equation*}
    C_{\mathrm{IL}}(h)
    \geq
    \frac{0+2}{2(2-0)}
    |h(2)-h(0)|
    =
    \frac{|h(2)|}{2}
    \geq
    \frac{a}{2}.
\end{equation*}
Thus, $C_K^\star \geq \frac{a}{2}.$

For $K\geq2$ and $b>0$, define
\begin{equation*}
    h_{K,b}(\lambda)
    \coloneqq
    -b\left[
        1-\left(1-\frac{\lambda}{2}\right)^K
    \right],
    \qquad
    \lambda\in[0,2].
\end{equation*}
Function $h_{K,b}$ is a polynomial of degree $K$, and it satisfies
\begin{equation*}
    h_{K,b}(0)=0,
    \qquad
    h_{K,b}(\lambda)\leq0
    \quad
    \text{for all }\lambda\in[0,2].
\end{equation*}

\noindent \textbf{Claim.} The integral Lipschitz constant of $h_{K,b}$ is
\begin{equation}
\label{eq: claim-CIL-hKb}
    C_{\mathrm{IL}}(h_{K,b})
    =
    \frac{b}{2}.
\end{equation}

\begin{proof}[Proof of the claim]
Let
\begin{equation*}
f_K(x)\coloneqq1-(1-x)^K.
\end{equation*}
Since
\begin{equation*}
h_{K,b}(\lambda)
=
-bf_K\left(\frac{\lambda}{2}\right),
\end{equation*}
setting
\begin{equation*}
x=\frac{\lambda_1}{2},
\qquad
y=\frac{\lambda_2}{2},
\end{equation*}
gives
\begin{equation*}
C_{\mathrm{IL}}(h_{K,b})
=
b
\sup_{0\leq x<y\leq1}
\frac{x+y}{2}
\frac{f_K(y)-f_K(x)}{y-x}.
\end{equation*}
For $0\leq x<y\leq1$, define
\begin{equation*}
Q_K(x,y)
\coloneqq
\frac{x+y}{2}
\frac{f_K(y)-f_K(x)}{y-x}.
\end{equation*}
We first want to show that for every $K\geq2$,
\begin{equation}
\label{eq: QK-half}
Q_K(x,y)\leq\frac{1}{2}.
\end{equation}
To see this, write
\begin{equation*}
u
\coloneqq
1-\frac{x+y}{2},
\qquad
v
\coloneqq
\frac{y-x}{2}.
\end{equation*}
Then
\begin{equation*}
0<u<1,
\qquad
0<v\leq\min\{u,1-u\},
\end{equation*}
and
\begin{equation*}
Q_K(x,y)
=
(1-u)
\frac{(u+v)^K-(u-v)^K}{2v}.
\end{equation*}
For fixed $u$,
\begin{equation*}
\frac{(u+v)^K-(u-v)^K}{2v}
=
\sum_{\substack{j=1\\ j\ \mathrm{odd}}}^{K}
\binom{K}{j}
u^{K-j}v^{j-1},
\end{equation*}
which is nondecreasing in $v$, since it contains only even powers of
$v$ with nonnegative coefficients.

If $0<u\leq1/2$, then $v\leq u$, and therefore
\begin{equation*}
\begin{aligned}
Q_K(x,y)
&\leq
(1-u)
\frac{(2u)^K}{2u}\\
&=
(1-u)2^{K-1}u^{K-1} \leq
\frac{1}{2},
\end{aligned}
\end{equation*}
where the last inequality follows because
$(1-u)u^{K-1}$ is nondecreasing on $(0,1/2]$ for $K\geq2$.

If instead $1/2<u<1$, then $v\leq1-u$, and therefore
\begin{equation*}
\begin{aligned}
Q_K(x,y)
&\leq
(1-u)
\frac{
1-(2u-1)^K
}{
2(1-u)
}\\
&=
\frac{1-(2u-1)^K}{2}
\leq
\frac{1}{2}.
\end{aligned}
\end{equation*}
Thus \eqref{eq: QK-half} holds for all $0\leq x<y\leq1$.
Moreover, $Q_K(0,1)=\frac{1}{2}$ and
therefore
\begin{equation*}
\sup_{0\leq x<y\leq1}Q_K(x,y)
=
\frac{1}{2},
\end{equation*}
which proves \eqref{eq: claim-CIL-hKb}.
\end{proof}

It remains to choose $b$ such that $h_{K,b}$ is feasible, i.e.,
\begin{equation*}
h_{K,b}(\lambda)
\leq
\frac{1}{2}\ln\left(\frac{\eta}{\lambda}\right),
\qquad
\lambda\in(0,2].
\end{equation*}
For $\lambda\in(0,\eta)$, the right-hand side is positive, while
$h_{K,b}(\lambda)\leq0$, so the constraint is automatically satisfied. For $\lambda\in[\eta,2]$, the smoothness
constraint can be written as
\begin{equation*}
-h_{K,b}(\lambda)
\geq
\frac{1}{2}\ln\left(\frac{\lambda}{\eta}\right)
=
a+\frac{1}{2}\ln\left(\frac{\lambda}{2}\right).
\end{equation*}
Using the definition of $h_{K,b}$, this gives
\begin{equation*}
b
\geq
\frac{
a+\frac{1}{2}\ln\left(\frac{\lambda}{2}\right)
}{
1-\left(1-\frac{\lambda}{2}\right)^K
}.
\end{equation*}
The smallest $b$ that guarantees feasibility for all
$\lambda\in[\eta,2]$ is
\begin{equation*}
b_K
=
\max_{\lambda\in[\eta,2]}
\frac{
a+\frac{1}{2}\ln\left(\frac{\lambda}{2}\right)
}{
1-\left(1-\frac{\lambda}{2}\right)^K
}.
\end{equation*}
Evaluating at $\lambda=2$ gives
$b_K\geq a.$ Moreover, for $\lambda\in[\eta,2]$,
\begin{equation*}
a+\frac{1}{2}\ln\left(\frac{\lambda}{2}\right)
\leq
a,
\end{equation*}
and
\begin{equation*}
1-\left(1-\frac{\lambda}{2}\right)^K
\geq
1-\left(1-\frac{\eta}{2}\right)^K
=
1-r^K.
\end{equation*}
Hence $b_K
\leq
\frac{a}{1-r^K}.$ Thus, $h_{K,b_K}$ is feasible, and by
\eqref{eq: claim-CIL-hKb},
\begin{equation*}
C_K^\star
\leq
C_{\mathrm{IL}}(h_{K,b_K})
=
\frac{b_K}{2}
\leq
\frac{a}{2(1-r^K)}.
\end{equation*}
Combining this with the lower bound $C_K^\star\geq a/2$ gives
\begin{equation*}
\frac{a}{2}
\leq
C_K^\star
\leq
\frac{a}{2(1-r^K)},
\end{equation*}
which completes the proof.
\end{proof}

%%%%%%%%%%%%%%%%%%%%%%%%%%%%%%%%%%%%%%%%%%%%%%%%%%%%%%%%%%
%
%                   Appendix G
%
%%%%%%%%%%%%%%%%%%%%%%%%%%%%%%%%%%%%%%%%%%%%%%%%%%%%%%%%%%

\section{Proof of Proposition \ref{prop: finite-order-optimality}}\label{proof: finite-order-optimality}
\begin{proof}
Evaluating \eqref{eq: b-hat} at $x=1$ gives
\begin{equation*}
b_K\geq a.
\end{equation*}
We derive a condition on $K$ ensuring that $b_K\leq a$. From \eqref{eq: b-hat}, it is sufficient that
\begin{equation*}
\frac{
a+\frac{1}{2}\ln x
}{
1-(1-x)^K
}
\leq
a,
\qquad
x\in[\eta/2,1].
\end{equation*}
Equivalently, $2a(1-x)^K
\leq
-\ln x.$ 
Using $-\ln x\geq1-x$, it is sufficient that
\begin{equation*}
2a(1-x)^{K-1}
\leq
1,
\qquad
x\in[\eta/2,1].
\end{equation*}
Since $1-x\leq1-\eta/2=r$, this is guaranteed by
$2ar^{K-1}\leq1.$ 
Taking logarithms and using $\ln r<0$ gives
\begin{equation*}
K
\geq
1+
\frac{\ln(2a)}{-\ln r}.
\end{equation*}
Thus, for integer $K\geq2$, it is sufficient that
\begin{equation*}
K
\geq
\max\left\{
2,\,
1+
\left\lceil
\frac{\ln(2a)}{-\ln r}
\right\rceil
\right\}
=
K_0.
\end{equation*}
Hence $b_K\leq a$. Together with $b_K\geq a$, this yields $b_K = a$, which completes the proof.
\end{proof}

\section{Topological Flow Matching}
\label{app:flow-matching}

We recall some results from~\cite{wyrwal2026topological} that characterize the conditional paths and vector fields of linear Gaussian reference processes, and specialize them to the polynomial graph filters considered in this work.

Consider the reference process SDE
\begin{equation*}
d\mathbf{y}_t = [\mathbf{A}\mathbf{y}_t+\boldsymbol{\alpha}_t]dt+\sigma\,d\mathbf{w}_t,
\end{equation*}
where $\mathbf{A}$ is a constant symmetric matrix and $\sigma>0$. As established in~\cite[Prop. 3]{wyrwal2026topological}, in the limit $\sigma\to0$, the bridge law conditioned on $(\mathbf{y}_0,\mathbf{y}_1)$ converges to a Dirac measure concentrated on its conditional mean,
\begin{equation}
\label{eq:app-bridge-mean}
\mathbf{m}_t^{\mathbf{y}_0,\mathbf{y}_1}
=
\mathbf{m}_{t|0}(\mathbf{y}_0)
+
\mathbf{R}_{t,1|0}
\mathbf{R}_{1,1|0}^{-1}
(\mathbf{y}_1-\mathbf{m}_{1|0}(\mathbf{y}_0)).
\end{equation}
Here, with $\mathbf{T}_{s,t}=e^{(t-s)\mathbf{A}}$ denoting the state-transition matrix, the conditional mean and normalized covariance of the reference process, conditioned on $\mathbf{y}_0$, are given by
\begin{equation*}
\mathbf{m}_{t|0}(\mathbf{y}_0)
=
\mathbf{T}_{0,t}\mathbf{y}_0
+
\int_0^t\mathbf{T}_{r,t}\boldsymbol{\alpha}_rdr,
\end{equation*}
and
\begin{equation*}
\mathbf{R}_{t,s|0}
=
\int_0^{\min(t,s)}
\mathbf{T}_{r,t}\mathbf{T}_{r,s}dr,
\end{equation*}
respectively. Furthermore, the conditional vector field in the zero-noise limit is given by~\cite[Prop. 2]{wyrwal2026topological}
\begin{equation}
\label{eq:app-limiting-control}
u_t^{\mathbf{y}_0,\mathbf{y}_1}
=
\mathbf{T}_{t,1}
\mathbf{R}_{1,1|0}^{-1}
(\mathbf{y}_1-\mathbf{m}_{1|0}(\mathbf{y}_0)).
\end{equation}

Regressing against the conditional vector field $u_t^{\mathbf{y}_0,\mathbf{y}_1}$ under any coupling $\gamma$ yields a valid flow-matching training objective, while the entropic optimal transport coupling~\cite[Prop. 4]{wyrwal2026topological} recovers the solution to the degenerate Schr\"odinger bridge problem~\cite[Prop. 1]{wyrwal2026topological}.\vspace{2pt}

\noindent\textbf{Polynomial graph filters.}
We specialize these results to $\mathbf{A}=\mathbf{H}(\mathbf{L})$, where
\begin{equation*}
\mathbf{H}(\mathbf{L})=\sum_{k=0}^{K}h_k\mathbf{L}^k,
\qquad
h(\lambda)=\sum_{k=0}^{K}h_k\lambda^k,
\end{equation*}
and set $\boldsymbol{\alpha}_t=\mathbf{0}$. Given the eigendecomposition $\mathbf{L}=\mathbf{V}\boldsymbol{\Lambda}\mathbf{V}^{\top}$, the graph filter $\mathbf{H}(\mathbf{L})$ shares the eigenvectors $\mathbf{V}$ of $\mathbf{L}$, with eigenvalues $\{h(\lambda_i)\}_{i=1}^N$. Defining $h_i=h(\lambda_i)$ and the GFT $\hat{\mathbf{y}}=\mathbf{V}^{\top}\mathbf{y}$, the state-transition matrix becomes
\begin{equation*}
\mathbf{T}_{s,t}
=
\mathbf{V}\operatorname{diag}(e^{h_i(t-s)})\mathbf{V}^{\top}.
\end{equation*}
Consequently, the normalized covariance is diagonal in the graph Fourier basis,
\begin{equation*}
\mathbf{V}^{\top}
\mathbf{R}_{t,s|0}
\mathbf{V}
=
\operatorname{diag}(q_i(t,s)),
\end{equation*}
where
\begin{equation*}
q_i(t,s)
=
\int_0^{\min(t,s)}e^{h_i(t+s-2r)}dr.
\end{equation*}
In particular,
\begin{equation*}
\mathbf{V}^{\top}
\mathbf{R}_{t,1|0}
\mathbf{R}_{1,1|0}^{-1}
\mathbf{V}
=
\operatorname{diag}\left(\frac{q_i(t,1)}{q_i(1,1)}\right)
=
\operatorname{diag}(c_{t,i}),
\end{equation*}
with
\begin{equation*}
c_{t,i}
=
\begin{cases}
e^{h_i(1-t)}\dfrac{e^{2h_i t}-1}{e^{2h_i}-1},
& h_i\neq0,\\
t, & h_i=0.
\end{cases}
\end{equation*}
Substituting into \eqref{eq:app-bridge-mean} yields the deterministic conditional path
\begin{equation}
\label{eq:app-spectral-path}
\hat{y}_{t,i}
=
e^{h_i t}\hat{y}_{0,i}
+
c_{t,i}(\hat{y}_{1,i}-e^{h_i}\hat{y}_{0,i}).
\end{equation}
Similarly, specializing \eqref{eq:app-limiting-control} gives the conditional vector field
\begin{equation}
\label{eq:app-spectral-control}
\hat{u}_{t,i}^{\mathbf{y}_0,\mathbf{y}_1}
=
a_{t,i}(\hat{y}_{1,i}-e^{h_i}\hat{y}_{0,i}),
\end{equation}
where
\begin{equation*}
a_{t,i}
=
\begin{cases}
e^{h_i(1-t)}\dfrac{2h_i}{e^{2h_i}-1},
& h_i\neq0,\\
1, & h_i=0.
\end{cases}
\end{equation*}
These expressions generalize the heat-equation conditional paths of~\cite{wyrwal2026topological}, recovered by setting $h_i=-\kappa\lambda_i$, to arbitrary polynomial graph filters. The conditional paths and vector fields derived here are transformed back to the vertex domain using the inverse GFT and employed to train the learned vector field $u_t^{\boldsymbol{\theta}}$, as tabulated in Algorithm~\ref{alg:topological-flow-matching}.